**Path distributions for describing eigenstates of the harmonic oscillator and other 1-dimensional problems**

Randall M. Feenstra
Dept. Physics, Carnegie Mellon University, Pittsburgh, PA 15213

## Abstract

The manner in which probability amplitudes of paths sum up to form wave functions of a harmonic oscillator, as well as other, simple 1-dimensional problems, is described. Using known, closed-form, path-based propagators for each problem, an integral expression is written that describes the wave function. This expression conventionally takes the form of an integral over initial locations of a particle, but it is re-expressed here in terms of a characteristic momentum associated with motion between the endpoints of a path. In this manner, the resulting expression can be analyzed using a generalization of stationary-phase analysis, leading to distributions of paths that exactly describe each eigenstate. These distributions are valid for all travel times, but when evaluated for long times they turn out to be real, non-negative functions of the characteristic momentum. For the harmonic oscillator in particular, a somewhat broad distribution is found, peaked at value of momentum that corresponds to a classical energy which in turn equals the energy eigenvalue for the state being described.

## I. Introduction

Teaching of quantum mechanics through an approach centered around Feynman path integrals [1,2,3,4,5] has been proposed by Taylor and others [6,7,8,9,10]. Although some undergraduate-level textbooks include a chapter on path integrals [11,12,13], this chapter often deals only with free-particle motion in Euclidean space. We are in full agreement that it is essential to treat that problem in detail. However, we also feel that there is considerable benefit in explaining the origin of wave functions *for all problems* in terms of paths. For this purpose, we describe in this work a method that utilizes closed-form expressions of propagators (solutions of the path integrals), analyzing those in the limit of long travel-time so as to deduce a distribution of paths that dominantly contribute to the motion within any given eigenstate.

The benefit of explaining eigenstates in terms of paths is that it builds upon the students' knowledge of the associated wave functions, while at the same time displaying certain dynamics of the particle that are apparent in the path-integral description but *not* in the time-dependent wave function itself. To illustrate this point, consider an experiment in which an electron, originating *e.g.* from a hot filament, is sent through an electron spectrometer [14] such that it exits this instrument being very nearly in an eigenstate of energy. Then, at some (large) distance from the spectrometer, the path along which the electron travels is crossed by a beam of different particles, such as high-energy photons or some other neutral particle. An array of electron detectors is placed around this crossing point, and occasionally a signal will be detected arising from a scattered electron.

What can be said concerning the dynamics of this scattered electron? Prior to the collision with the neutral particle, the electron was in an eigenstate of energy (and momentum). Consequently, in accordance with the position-momentum uncertainty relation, we cannot say anything whatsoever about where the electron is located (*i.e.* all locations are equally probable). However, if we end up detecting the electron when it is scattered at the crossing point with the neutral beam, we *can* draw a limited conclusion about how the electron traveled to that point. To achieve this, as detailed in Section II, we start with the well-known expression that relates the wave function value at the final location to the closed-form, path-based expression of the propagator, written in terms of an integral over all initial locations. We rewrite that expression in terms of a variable that is the classical momentum associated with the initial and final locations of the travel, for a given travel time $T$. This expression can then be analyzed in the limit of $T \to \infty$, using the method of stationary phase. The result is that, for some large time $T$ prior to when we detected the electron, it was most probably, or *dominantly* (in the sense of stationary phase [15,16]) at a location which is a distance $\hbar kT/M$ back towards the spectrometer from where it was detected, where $M$ is the particle mass and $\hbar k$ is the momentum eigenvalue for the state that the electron was prepared in.

This conclusion concerning the distance traveled by the electron is quite obvious; we are only saying that the classical momentum, given by

$$p_{\mathrm{c}} = \frac{M(x_{\mathrm{f}} - x_0)}{T} \tag{1}$$

where $x_0$ and $x_{\mathrm{f}}$ are the initial and final locations of the particle, respectively, is nearly equal (in the sense of stationary phase) to the momentum eigenvalue $\hbar k$ for the state that we are considering. If this is the only conclusion that we can draw in this work, then it would be scarcely worth dwelling upon. However, if we consider motion in a harmonic oscillator then it turns out that conventional stationary-phase analysis is insufficient to handle the problem, since there is not just a single dominant path, nor a few of them, but there is a continuous distribution of paths that must be included in order to describe the motion. To obtain this distribution, we must employ a generalization of the stationary-phase method.

For situations other than a free particle, our analysis still proceeds by utilizing a path-based expression for the wave function, but we then rewrite that using a different momentum than in Eq. (1). We will refer to the new integration variable, in general, as a *characteristic* momentum, denoted $p_{\mathrm{c}}$ for linear momentum or $L_{\mathrm{c}}$ for angular momentum. This value characterizes a path (or more precisely, a set of paths, each producing the same value of characteristic momentum). Relationships analogous to Eq. (1) that relate the endpoints of the path to the characteristic momentum will be provided for all situations that we consider. Examples include motion around a circle, for which the definition of characteristic momentum includes a winding number, as well as the harmonic oscillator, for which the characteristic momentum is taken to be the *maximum* momentum along a classical path (occurring when the particle is at a minimum of the potential).

We point out that the $T \to \infty$ analysis that we perform is actually addressing a significantly different question than what is generally discussed using the classical limit of $\hbar \to 0$. With the latter limit, one is trying to understand why a particular path is favored from amongst the very large set of all possible paths that connect initial and final space-time locations. In contrast, the starting point for our problem, *i.e.* the closed-form propagator, has already dealt with all possible paths that differ from whatever *small set* of paths are explicitly used for writing the probability amplitude terms of the propagator. For example, semiclassical propagators are formed by just using classical trajectories to write the probability amplitude terms. In that case, prefactors of those terms are given by a fluctuation factor – the square-root of a Van Vleck-Pauli-Morette (VPM) determinant [3,4,17,18,19] which describes the influence of all paths other than the classical trajectory – times a homotopy factor if needed [20,21]. Given this sort of closed-form propagator, the question we are then addressing in our work is what combination of paths of this *small set* is needed in order to form the wave function of a given eigenstate.

In Section II we describe our method for determining the distribution of paths that describe a wave function, dealing with free-particle motion in 1-dimensional Euclidean space, $\mathbb{R}^1$. In Section III we provide results for motion around a circle (a space denoted by $S^1$)*, reflection from a hard wall, and motion in an infinite square-well. The various expressions and manipulations involved in these cases, aside from the final step in determining the path distributions, differ little from prior work [3,4,16,17,22,23,24,25,26], but nonetheless we include a brief description of those analyses. In Section IV we derive path distributions for the harmonic oscillator, and in Section V we compare those results with other sorts of distributions associated with those eigenstates. We also briefly discuss how path distributions may (or may not) be obtained for various other situations. Our work is summarized in Section VI.

## II. Method

### A. Propagators utilizing characteristic momentum rather than initial position

We consider motion in 1 spatial dimension. Starting from an integral that describes the wave function and employs as its integration variable the initial position of the particle, path distributions are obtained using a three-step procedure:
(i) transform this spatial integral, if it extends over a finite region of space, into one that extends over an infinite region (this transformation of the integral, in many cases, embodies the important concept of a "covering space" and the homotopic relationship between that space and the one which the spatial integral first extends over [3,26,27]);
(ii) rewrite the integral in terms of a characteristic momentum, which is a classical momentum that relates to the endpoints of the paths (the paths themselves are not required to be classical trajectories, only their endpoints need be connected by a classical, characteristic momentum);
(iii) further rewrite the integral as a phasor sum, and evaluate that in the limit of long travel-time using an expression that is a generalization of the method of stationary phase, applicable to integrands of arbitrary complexity.

Our starting point is the fundamental equation that determines the evolution of a quantum state $\Psi(x,t)$ from some initial time $t_0$ to a final time $t_{\mathrm{f}}$, [1,2]

$$\Psi(x_{\mathrm{f}},t_{\mathrm{f}}) = \int_{\Omega} \Psi(x_0,t_0)\, K(x_0,t_0,x_{\mathrm{f}},t_{\mathrm{f}})\, dx_0 \ , \tag{2}$$

where $K(x_0,t_0,x_{\mathrm{f}},t_{\mathrm{f}})$ is the propagator (kernel) and $\Omega$ refers to the extent of the space that we are considering. The goal of our analysis is to describe energy eigenstates in a situation with a time-independent potential, so that the time evolution in Eq. (2) is given by $\Psi_j(x,t) = \psi_j(x) e^{-iE_j t/\hbar}$ with $j$ labeling the particular eigenstate. Inserting this form into the equation, and moving (for convenience) the $e^{-iE_j t_{\mathrm{f}}/\hbar}$ term from the left-hand side to the right, yields

$$\psi_j(x_{\mathrm{f}}) = e^{iE_j T/\hbar} \int_{\Omega} \psi_j(x_0)\, K(x_0,x_{\mathrm{f}},T)\, dx_0 \tag{3}$$

with the travel time given by $T \equiv t_{\mathrm{f}} - t_0$ and with the propagator now relabeled as $K(x_0,x_{\mathrm{f}},T)$ since its time dependence involves only $T$. In our analysis, we will insert a path-based form of the propagator into the integral on the right-hand side of this equation, *i.e.* written in terms of paths that extend between the initial position $x_0$ of the particle and its final position $x_\mathrm{f}$. We utilize known, closed-form expressions for the propagators, written in terms of a relatively small number of paths (a finite or countably infinite number, as opposed to the uncountably infinite number that would occur in a time-sliced expression).

---

* The fact that we use different types of symbols to refer to Euclidean space or an *n*-sphere (double-struck and italic, respectively) is of no significance; we are simply employing the most commonly used symbols for these spaces.

The first step of our method is to rewrite Eq. (3), if needed, in terms of an equivalent problem having initial positions $\tilde{x}_0$ that exist over an *extended* space, with $-\infty < \tilde{x}_0 < \infty$. The positions $\tilde{x}_0$ of the equivalent problem are not necessarily equal to $x_0$ since, as we will demonstrate, there can be more than value of $\tilde{x}_0$ that extends (along a path with endpoints connected by the characteristic momentum $p_\mathrm{c}$) to a given final location. However, for these final locations we can, without loss of generality, take $\tilde{x}_\mathrm{f} = x_\mathrm{f}$. Thus, we have for this equivalent problem

$$\psi_j(x_\mathrm{f}) = e^{iE_jT/\hbar} \int_{-\infty}^{\infty} \psi_j(\tilde{x}_0)\, \tilde{K}(\tilde{x}_0, x_\mathrm{f}, T)\, d\tilde{x}_0 \ . \qquad (4)$$

This rewriting of Eq. (3) is accomplished, *e.g.* for motion in $S^1$ or in an infinite square-well, by combining the original integral over $\Omega$ with a sum over an integer within the propagator [16,17,22, 23,24,25,26]. The resulting form of the propagator in Eq. (4) thus changes (it no longer has the sum over the integer), as indicated by the new function $\tilde{K}$. For the case of the harmonic oscillator, this step in the analysis is not necessary since its wave function extends over $-\infty < x_0 < \infty$ to begin with.

The second step is to change the variable of integration from the position $\tilde{x}_0$ to a characteristic linear momentum $p_\mathrm{c}$ (or angular momentum $L_\mathrm{c}$), as already introduced in Eq. (1). This momentum is determined by the endpoints of the travel. Importantly, it must be a *signed* quantity, with values that extend from $-\infty$ to $+\infty$, such that the third step of the analysis is enabled. Utilizing this variable, then the initial position $\tilde{x}_0$ is a function of $p_\mathrm{c}$, so Eq. (4) can be written as

$$\psi_j(x_\mathrm{f}) = e^{iE_jT/\hbar} \int_{-\infty}^{\infty} \psi_j(\tilde{x}_0(p_\mathrm{c}))\, \breve{K}(p_\mathrm{c}, x_\mathrm{f}, T)\, dp_\mathrm{c} \qquad (5)$$

where we again have a new form for the propagator, $\breve{K}(p_\mathrm{c}, x_\mathrm{f}, T)$. For example, for free-particle motion in $\mathbb{R}^1$ we have Eq. (1) for $p_\mathrm{c}$, so that $\tilde{x}_0(p_\mathrm{c}) = x_0(p_\mathrm{c}) = x_\mathrm{f} - p_\mathrm{c}T/M$. Hence, $\tilde{K}(\tilde{x}_0(p_\mathrm{c}), x_\mathrm{f}, T) = \tilde{K}(x_\mathrm{f} - p_\mathrm{c}T/M\,, x_\mathrm{f}, T)$, and this function will combine with a $d\tilde{x}_0/dp_\mathrm{c}$ term to form $\breve{K}(p_\mathrm{c}, x_\mathrm{f}, T)$ in Eq. (5). However, in general the relationship between $\tilde{x}_0$ and $p_\mathrm{c}$ is more complicated than that. For example, for reflection off of a wall or motion in an infinite square-well, the particle can approach the final position from the $x < x_\mathrm{f}$ or $x > x_\mathrm{f}$ (corresponding to $p_\mathrm{c} > 0$ or $p_\mathrm{c} < 0$, respectively), with different forms of $\tilde{x}_0(p_\mathrm{c})$ in the two cases, as described in Section III. For a harmonic oscillator this step of transforming from a spatial variable to the momentum variable is similarly nontrivial, as discussed in Section IV.

The reason that we rewrite Eq. (4) as Eq. (5) is that this form enables evaluation of the integrand utilizing the method of stationary phase (or our generalization of that method), in the limit of long travel-time $T \to \infty$. It is natural that we end up working in this limit, since we are describing motion within a given eigenstate of energy, and in accordance with the energy-time uncertainty relation it is only in the long-time limit that it makes sense to say that a particle is in a specific eigenstate. Equation (5) is recognized as providing a possible means of evaluating the contribution that paths with specific value of $p_c$ makes towards forming the wave function of a given eigenstate, *i.e.* utilizing the integrand

$$e^{iE_jT/\hbar}\,\psi_j(\tilde{x}_0(p_c))\,\breve{K}(p_c,x_f,T) \tag{6}$$

as a possible measure of that, per unit $p_c$-value. This contribution would correspond to a particular value of $x_f$ and a particular travel time $T$, although if we were to perform suitable averaging over those quantities (in a manner that we will specify in Section II(C)), then it would seem at least possible that one might obtain a distribution that is indicative of how the paths contribute towards the wave function.

However, Eq. (6) as it stands (with or without averaging over $x_f$ and $T$) turns out, in many cases, to work very poorly for providing a measure of the contribution that paths makes towards forming the wave function. In a rigorous, mathematical sense [5], only a real-valued, non-negative quantity is regarded as providing a well-defined *measure* within an integrand, and Eq. (6) as it stands does not provide that. Rather, there are complex-valued oscillations that can occur in Eq. (6) as a function of $p_c$, and these oscillations become more and more rapid as $T$ increases. Nevertheless, when we integrate Eq. (6) (as in Eq. (5)) to form the wave function, it turns out in some cases that only paths which are centered narrowly about a single value of $p_c$ (or a number of well separated values) end up contributing significantly to the integral. This is the usual situation for integrals that can be evaluated by the method of stationary phase [15,16].

In this manner, when Eq. (6) has just one or a discrete set of stationary phases as a function of $p_c$, it is possible to evaluate the contribution that paths make to the wave function on the left-hand side of Eq. (5): The *dominant* paths, in the limit of $T\to\infty$, comes just from the $p_c$-values associated with the stationary phases [15,16]. However, not all of the problems that we analyze in this work can be understood in that manner. In particular, the harmonic oscillator is found to require a continuous (connected) set of $p_c$-values in order to evaluate the integral of Eq. (5), even in the $T\to\infty$ limit. We therefore require some other method for evaluating the respective contributions of the integrand (the paths) for these situations. In an effort to obtain a suitable measure of these contributions, we integrate Eq. (6) over a $2\sqrt{\hbar M/T}$-wide range of $p_c$ values.

Utilizing this average, we find that Eq. (5) can be expressed as

$$\psi_j(x_f)=\frac{e^{iE_jT/\hbar}}{2\sqrt{\hbar M/T}}\int_{-\infty}^{\infty}dp_c\int_{p_c-\sqrt{\hbar M/T}}^{p_c+\sqrt{\hbar M/T}}\psi_j(\tilde{x}_0(p'))\,\breve{K}(p',x_f,T)\,dp'\,. \tag{7}$$

It is by no means obvious that the right-hand sides of Eqs. (5) and (7) are equal, but we find that they indeed are, at least for the types of problems that we consider here. As we will illustrate in Section II(B), the integrals in Eqs. (5) and (7) can be conveniently viewed on a phasor diagram. For the problems considered here, all values on this phasor diagram are finite, and furthermore, the integral over $p_c$ (as in Eqs. (5) and (7)) has negligible contributions from $p_c$ values of very large magnitude. By rewriting Eq. (5) in the form of Eq. (7), we are changing the manner in which we sum up this phasor diagram, either as a continuous sum (as in Eq. (5)), or as a sum of segments and then averaging over the starting value that is used to define the segments (as in Eq. (7)). Either way, the same result is produced. Moreover, with the average over the $2\sqrt{\hbar M/T}$-wide range of $p_c$ values in the integrand of Eq. (7), a greatly improved measure is obtained of how paths contribute towards forming the wave function.

Proceeding with a derivation of Eq. (7), we first re-express Eq. (5) as

$$\psi_j(x_{\rm f}) = e^{iE_jT/\hbar} \sum_{\nu=-\infty}^{\infty} \int_{p_\nu-\sqrt{\hbar M/T}}^{p_\nu+\sqrt{\hbar M/T}} \psi_j(\tilde{x}_0(p_{\rm c}))\, \breve{K}(p_{\rm c}, x_{\rm f}, T)\, dp_{\rm c} \tag{8a}$$

with

$$p_\nu = 2\nu \sqrt{\frac{\hbar M}{T}} + \delta p \tag{8b}$$

where $\delta p$ is any finite value of momentum. The presence of $\delta p$ here is a consequence of the fact that the zero of momentum utilized for the $p_\nu$ values in Eq. (8) is arbitrary (since the integral of Eq. (5) extends from $-\infty$ to $+\infty$, and we are assuming that the contribution to the integral from $p_{\rm c}$ values of very large magnitude is negligible). If we average over values of $\delta p$ extending from 0 to $2\sqrt{\hbar M/T}$, Eq. (8a) becomes

$$\psi_j(x_{\rm f}) = \int_{-\infty}^{\infty} \frac{\Delta \mathcal{I}_j(p_{\rm c}, x_{\rm f}, T)}{2\sqrt{\hbar M/T}}\, dp_{\rm c} \tag{9a}$$

where we define both

$$\Delta \mathcal{I}_j(p_{\rm c}, x_{\rm f}, T) \equiv \mathcal{I}_j\left(p_{\rm c} + \sqrt{\hbar M/T}, x_{\rm f}, T\right) - \mathcal{I}_j\left(p_{\rm c} - \sqrt{\hbar M/T}, x_{\rm f}, T\right) \tag{9b}$$

and

$$\mathcal{I}_j(p_{\rm c}, x_{\rm f}, T) \equiv e^{iE_jT/\hbar} \int_{-\infty}^{p_{\rm c}} \psi_j\big(\tilde{x}_0(p')\big)\, \breve{K}(p', x_{\rm f}, T)\, dp' \quad . \tag{9c}$$

Equation (9a) is seen to be identical to Eq. (7). This equality is illustrated graphically in Section II(B), where a plot of $\mathcal{I}_j(p_{\rm c}, x_{\rm f}, T)$ as a function of $p_{\rm c}$ is shown to yield a familiar phasor diagram and associated stationary-phase features. The quantity $\Delta \mathcal{I}_j(p_{\rm c}, x_{\rm f}, T)/2\sqrt{\hbar M/T}$, evaluated for $T \to \infty$ and including averaging over $x_{\rm f}$ (and possibly $T$) as specified in Section II(C), turns out to provide a rigorous measure of the contribution that paths make towards a given wave function; evaluation of this measure constitutes the third and final step of our method.

### B. Illustrative example – motion in $\mathbb{R}^1$

To understand how the oscillatory nature of Eq. (6) can impact its integral, Eq. (5), let us consider the problem of free-particle motion in $\mathbb{R}^1$. The propagator for this motion is [2,3,4]

$$K^{(\mathbb{R}^1)}(x_0, x_{\rm f}, T) = \sqrt{\frac{M}{2\pi i \hbar T}}\; e^{iM(x_{\rm f}-x_0)^2/2\hbar T} \tag{10}$$

with eigenvalues $E_k = \hbar^2 k^2/2M$ for all real values of $k$, and wave functions $\psi_k(x) = e^{ikx}$ (a normalization factor such as $1/\sqrt{2\pi}$ can be included here, with no impact on the resulting path distribution, but our expressions are slightly simpler without that term so we do not include it). Inserting this propagator into Eq. (4) (with $\tilde{x}_0 \equiv x_0$ since $x_0$ varies between $\pm\infty$),

$$e^{ikx_\mathrm{f}} = e^{i\hbar k^2 T/2M} \sqrt{\frac{M}{2\pi i\hbar T}} \int_{-\infty}^{\infty} e^{ikx_0}\, e^{iM(x_\mathrm{f}-x_0)^2/2\hbar T}\, dx_0 \,. \tag{11}$$

Changing the variable of integration to be $p_\mathrm{c}$ of Eq. (1), we obtain

$$e^{ikx_\mathrm{f}} = e^{ikx_\mathrm{f}} \sqrt{\frac{T}{2\pi i\hbar M}} \int_{-\infty}^{\infty} e^{i(p_\mathrm{c}-\hbar k)^2 T/2\hbar M}\, dp_\mathrm{c} \,. \tag{12}$$

Of course, the integral on the right-hand side of Eq. (12) can be trivially evaluated as a Gaussian integral, thereby demonstrating the equality of the two sides of the expression. Nevertheless, our goal is to evaluate the contribution that various paths make towards forming the wave function, so we rewrite the integral in a form suitable for that. However, before doing so, we pause to point out an aspect of Eq. (12) that may be a distraction to readers, namely, the presence of the wave function evaluated at the final location, $e^{ikx_\mathrm{f}}$, on *both* sides of the equation. We are attempting to explain this $e^{ikx_\mathrm{f}}$ term on the left side in terms of the expression on the right, but there we also have an explicit $e^{ikx_\mathrm{f}}$ term. The presence of this term on the right can be somewhat distracting, to the point of perhaps suggesting some sort of manifest triviality (or even worse, a circular reasoning) in our analysis. We urge the reader not to be distracted by the presence of this term on the right. In more complicated situations, such as the harmonic oscillator discussed in Section IV or the problem of angular momentum discussed elsewhere [28], an analogous integral as in Eq. (12) is constructed (with general form as shown in Eq. (5)) but in those cases a final wave function term does *not* explicitly occur on the right-hand side. For the case of Eq. (12), this term appears (*i.e.* it was able to be factored out of the integral) due to the extreme simplicity of the situation being considered. However, for all cases, our goal is to evaluate the integral in such a way so as to deduce the contributions that various paths make towards forming the wave function on the left-hand side. Hence, for Eq. (12) in particular, we wish to determine the paths that dominantly determine a value of $\sqrt{2\pi i\hbar M/T}$ for the integral there, such that the two sides of the expression are equal.

Considering evaluation of the right-hand side of Eq. (12) with this goal in mind, it is clear that the method of stationary-phase [15,16] can be applied to that integral (taking a derivative of the exponent in the integrand and setting that equal to zero yields dominant values of $p_\mathrm{c}$ near $\hbar k$, with these dominant paths determining the value of $\sqrt{2\pi i\hbar M/T}$ for the integral, in the limit $T \to \infty$). However, we desire here a more general method, one that can be applied to *any* integrand, *i.e.* even if its form is too complicated to permit isolating an exponent and taking a derivative of that. Again, we are not so interested in the value of the integral (which we already know based upon the left-hand side of the expression), but rather, we wish to deduce the paths that dominantly determine the value of the integral.

For this purpose, we start by taking the right-hand side of Eq. (12) but utilizing an upper integration limit of $p_\mathrm{c}$, thereby yielding $\mathcal{I}_k$ of Eq. (9c),

$$\mathcal{I}_k(p_\mathrm{c}, x_\mathrm{f}, T) = e^{ikx_\mathrm{f}} \sqrt{\frac{T}{2\pi i\hbar M}} \int_{-\infty}^{p_\mathrm{c}} e^{i(p'-\hbar k)^2 T/2\hbar M}\, dp' \,. \tag{13}$$

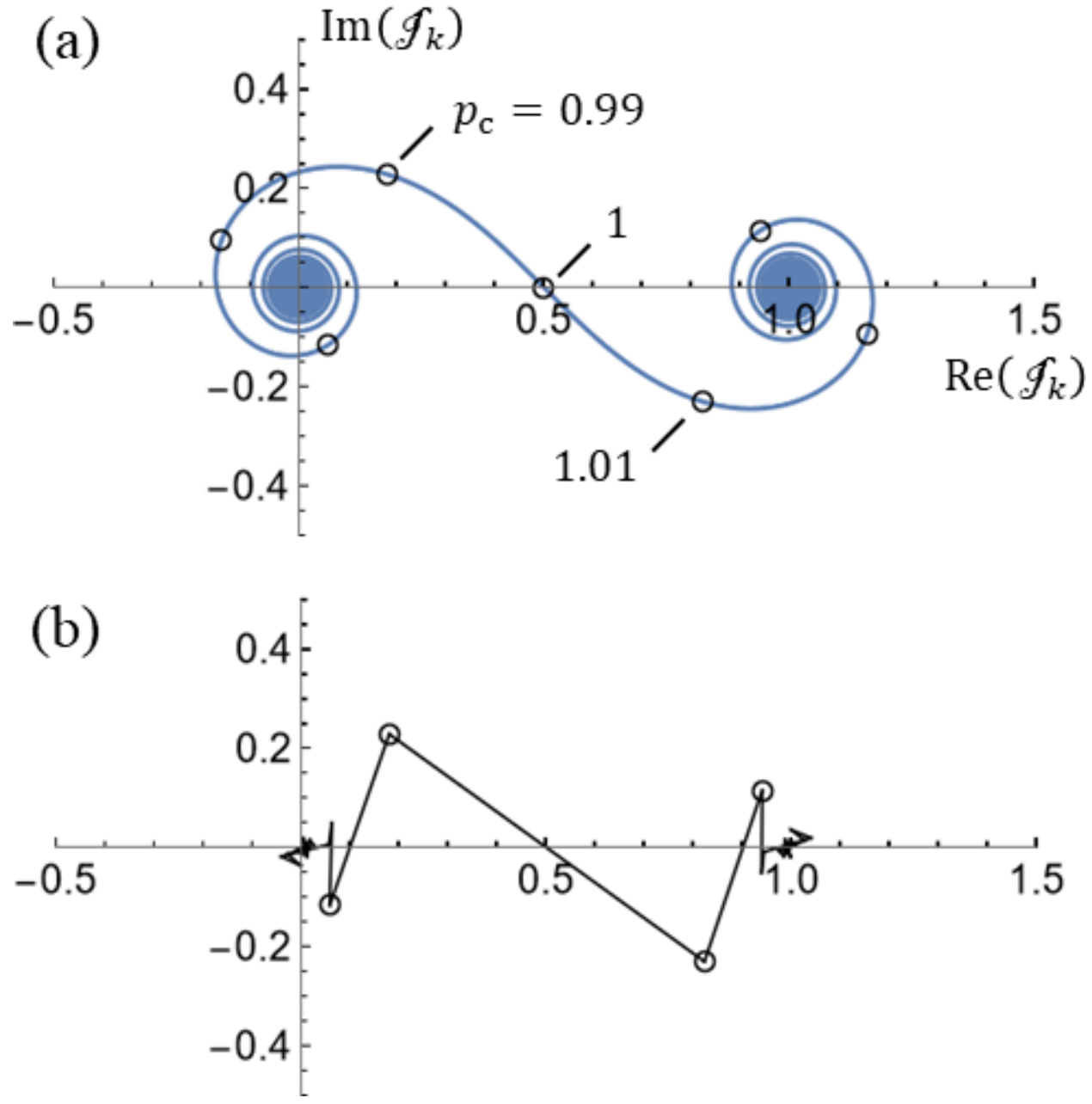

FIG 1. (a) Phasor diagram of $\mathcal{J}_k(p_c, x_f, T)$ from Eq. (13), as a function of $p_c$ varying between $-\infty$ and $+\infty$, for an eigenstate with $k = 1$ at a travel time of $T = 10^4$. The black circles indicate $p_c$ values of 0.97, 0.98, 0.99, 1, 1.01, 1.02, and 1.03 (with the middle three values labeled). Units of $\hbar = M = 1$ are used. (b) Decomposition of the curve in (a) into line segments that join a set of $p_c$ values spaced by 0.02.

Values of this quantity are shown on the complex plane in Fig. 1(a), using $\hbar = M = 1$, an eigenvalue of $k = 1$, a final position of $x_f = 0$, and a travel time of $T = 10^4$. It is clear from this phasor diagram that values of $p_c$ which are very close to $\hbar k$ (within a few multiples of $\sqrt{\hbar M/T} = 0.01$) are dominant in forming the integral of Eq. (13). Thus, by the usual arguments of the method of stationary phase [15,16], applied here in the $T \to \infty$ limit, we can view the wave function as being essentially determined by these paths whose endpoints are associated with a $p_c$ value that is very close to $\hbar k$. Figure 1(b) shows line segments, corresponding to $\Delta\mathcal{J}_j(p_c, x_f, T)$ of Eq. (9b), that extend between a specific set of $p_c$ values spaced by 0.02. These line segments sum up to yield the same result as for the continuous phasor sum of Fig. 1(a), and this equality of the two sums holds even if a shift ($\delta p$ in Eq. (8b)) is made in the $p_c$ values that the line segments extend between.

Let us now consider, following the discussion surrounding Eq. (6), whether or not a useful measure of the contribution that a path makes towards forming a given wave function is provided by the integrand of Eq. (12) (together with the factors multiplying that integral),

$$e^{ikx_f} \sqrt{\frac{T}{2\pi i\hbar M}}\, e^{i(p_c-\hbar k)^2 T/2\hbar M} \quad . \tag{14}$$

In Fig. 2(a) we plot this quantity, for the same parameters as in Fig. 1. We find strongly oscillatory behavior, as expected in integrands such as this that are dominated by exponents which have a stationary phase [15,16].

To form a much improved measure of how paths contribute towards wave functions, we then employ an average of the function in Eq. (14), corresponding to $\Delta\mathcal{J}_j(p_c, x_f, T)/2\sqrt{\hbar M/T}$ as defined by Eqs. (9b) and (9c),

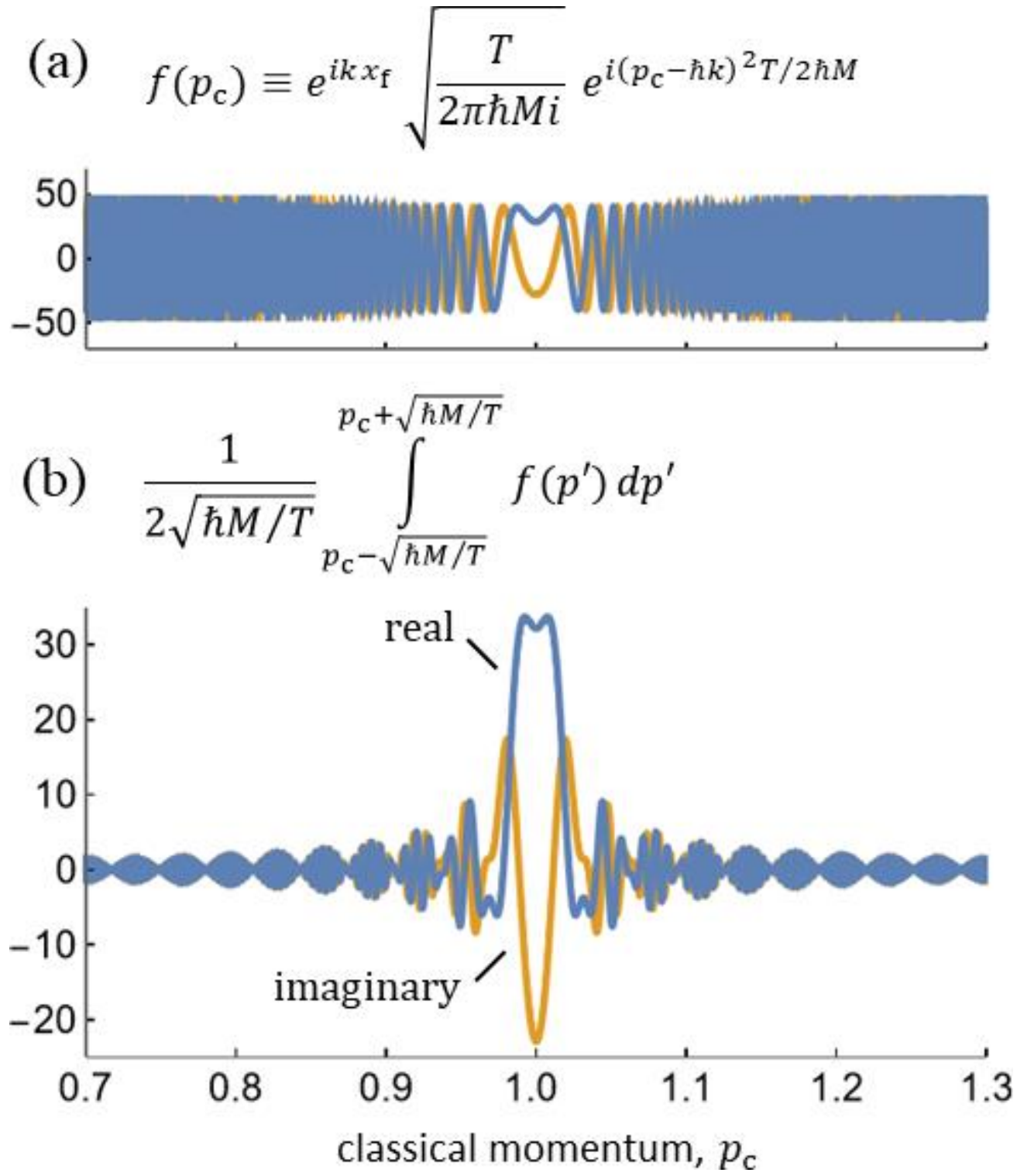

FIG 2. (a) Plot of Eq. (14), defined here as the function $f(p_\mathrm{c})$. (b) Plot of Eq. (15), which is an average of $f(p_\mathrm{c})$ over a $p_\mathrm{c}$ range with width $2\sqrt{\hbar M/T}$. The same parameter values as in Fig. 1 are used, with units of $\hbar = M = 1$. Real and imaginary parts of the plotted quantities are indicated.

$$\frac{\Delta\mathcal{J}_k(p_\mathrm{c}, x_\mathrm{f}, T)}{2\sqrt{\hbar M/T}} = \frac{e^{ikx_\mathrm{f}}}{2\sqrt{\hbar M/T}} \sqrt{\frac{T}{2\pi i\hbar M}} \int_{p_\mathrm{c}-\sqrt{\hbar M/T}}^{p_\mathrm{c}+\sqrt{\hbar M/T}} e^{i(p'-\hbar k)^2 T/2\hbar M}\, dp' \,. \tag{15}$$

A plot of this quantity (which can be easily evaluated in terms of imaginary error functions) is shown in Fig. 2(b), obtained using computer code listed in the Appendix and available at https://www.cmu.edu/physics/stm/publ/136/. A peak centered at the $p_\mathrm{c}$ value of 1 (which corresponds to $\hbar k$ in this example) is apparent. This peak becomes increasingly narrow as $T$ increases, with the oscillatory structure seen on either side of the peak also becoming closer and closer to the peak. Hence, the real part of Eq. (15) in the $T \to \infty$ limit behaves equivalently to a Dirac delta function when utilized within an integral (*i.e.* in just the same sense that $\sin(\lambda x)/\pi x$ approaches a delta function in the limit $\lambda \to \infty$ [29]). Concerning the imaginary part of Eq. (15), as is apparent in Fig. 2(b) this takes the form of a second derivative of a Gaussian, which in the $T \to \infty$ limit behaves equivalently to a second derivative of a Dirac delta function. This second derivative, when integrated over $p_\mathrm{c}$ (and with no other function in the integrand multiplying the second derivative), makes zero contribution to the value of the integral. Thus, we can say that Eq. (15) behaves (within an integral) like $\exp(ikx_\mathrm{f})\,\delta(p_\mathrm{c} - \hbar k)$ for $T \to \infty$. When this is integrated over $p_\mathrm{c}$ as in Eq. (9a), we obtain the wave function at the final location, $\exp(ikx_\mathrm{f})$.

To further investigate this result, it is informative to step back to Eq. (11). Since we found that the integral in Eq. (12) has, for $T \to \infty$, significant contributions only from $p_\mathrm{c}$ values very near $\hbar k$, it would seem that the integral of Eq. (11) should similarly be dominated by some range of paths centered about $x_0 = x_\mathrm{f} - \hbar kT/M$. Even though the method of stationary phase cannot be applied directly to Eq. (11), since there is no large parameter that multiplies the phase within the exponent, we nevertheless note that a derivative of the exponent (without any multiplier) with respect to $x_0$ does indeed yield a stationary phase at $x_0 = x_\mathrm{f} - \hbar k T/M$. In fact, a phasor diagram of Eq. (11) (*i.e.* writing it with an upper limit in the integral of $x_0$ and then evaluating it as a function of $x_0$) yields an identical plot as Fig. 1(a), so long as we linearly rescale the variable of the plot from $p_\mathrm{c}$ in Fig. 1(a) to $x_0 = x_\mathrm{f} - \hbar k T/M$. Of course, the significant values of $x_0$ in determining the value of the integral in Eq. (11) are negative and with large magnitude (assuming some finite value for $x_\mathrm{f}$), since $T \to \infty$. Nevertheless, considering a range of $x_0$ values that scales like $\sqrt{\hbar T/M}$ relative to where the stationary phase occurs, then the $x_0$-integral of Eq. (11) can be understood in just the same manner as the $p_\mathrm{c}$-integral of Eq. (12).

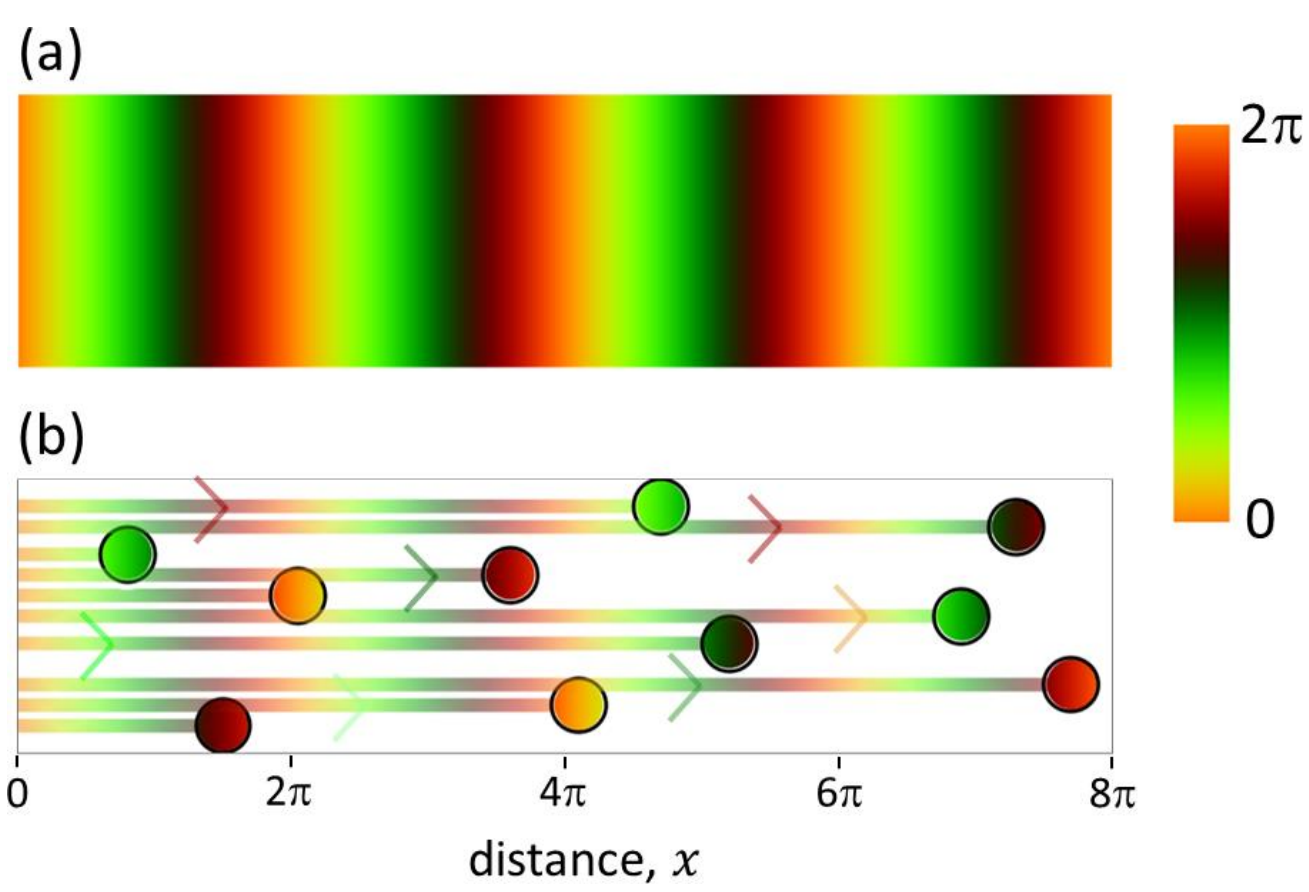

FIG 3. (a) Wave function for free-particle motion in $\mathbb{R}^1$, with phase indicated by the color scale shown on the right-hand side. (b) Dominant paths for the motion. Final locations are indicted by circles, filled by a color indicating the phase. (The vertical extent of the plots has no physical meaning; it is used only to provide space to show the various paths).

With this property of Eq. (11) in mind, we can then construct a very simple illustration of our result of a dominant path with characteristic momentum $p_\mathrm{c} = \hbar k$. Figure 3(a) pictures the wave function $\psi_k(x_\mathrm{f}) = e^{ikx_\mathrm{f}}$, with its phase shown by the color-scale indicated. We take $k = 1$, using dimensionless quantities with $\hbar = M = 1$ throughout. Then, in Fig. 3(b) we picture dominant paths, starting at various initial locations far off the left-hand edge of the plot (since the travel time $T$ is large), and extending to the final locations as shown by the filled circles. With $x_\mathrm{f} = x_0 + p_\mathrm{c} T/M$ and $p_\mathrm{c} = \hbar k$, the dominant paths are shown in accordance with the trivial decomposition of the wave function,

$$e^{ikx_\mathrm{f}} = e^{ikx_0}\, e^{ikp_\mathrm{c}T/M} \tag{16a}$$

$$= e^{i\hbar k^2 T/2M}\, e^{ikx_0}\, e^{i(p_\mathrm{c})^2 T/2\hbar M}\,. \tag{16b}$$

The wave function at its final location is thusly written as a product of a $e^{i\hbar k^2 T/2M}$ term from the energy eigenvalue (the leading term in Eqs. (4) and (11)) together with a wave function at the initial location $e^{ikx_0}$ and a probability amplitude term $e^{i(p_\mathrm{c})^2 T/2\hbar M}$ (the two terms within the integrals of Eqs. (4) and (11)). This decomposition makes sense because, as just argued, the integral of Eq. (11) does indeed have significant contributions only from some range of $x_0$ values centered around $x_\mathrm{f} - \hbar k T/M$. In Fig. 3(b) we plot the phase of the product in Eq. (16b), further allowing the travel time to vary as $0 \le t \le T$ and plotting the location of the end-point of the path at an $x$ value of $x = x_0 + p_\mathrm{c} t/M$. The phases as seen in Figs. 3(a) and 3(b) are obviously in agreement.

This is the type of decomposition of the wave function into its dominant paths that we are striving to achieve for all the problems discussed in this work. The result for motion in $\mathbb{R}^1$ as shown in Fig. 3 is exceptionally simple, since not only are the phases at the *final locations* in Fig. 3(b) in agreement with those of Fig. 3(a), but actually the phases *over the entire paths* shown there are in agreement with the wave function. However, that situation is specific only to problems in which there is just a single type of dominant path (as further discussed at the end of Section III(B)), as occurs *e.g.* for motion in $\mathbb{R}^1$ or $S^1$. The picture of the path-based motion shown in Fig. 3(b) is in good accord with the experimental detection of particles discussed in the third paragraph of Section I.

Let us now consider what value of $T$ is needed in order to reasonably achieve a $T \to \infty$ limit. Considering $\hbar = M = 1$ Eq. (13), but varying $k$, we find that $T$ must be scaled by $k^{-2}$ such that *increments* in $p_{\mathrm{c}}$ produce the same spacings in a phasor plot as found in Fig. 1(a). With $T \gg 1$ in Fig. 1(a) such that the $T \to \infty$ limit is achieved, we thus conclude that $T \gg k^{-2}$ in general. Explicitly including $\hbar$ and $M$ values, then the limit is achieved when $T \gg \hbar M/(\hbar k)^2 = \hbar/2E_k$, which yields 0.33 fs for electrons with energy eigenvalue of $E_k = 1$ eV. During this time, the electron travels $\hbar/\sqrt{2ME_k} \approx 2.0$ Å. The large $T$ limit is thus readily achieved (at least on the scale of atoms), unless $E_k \ll 1$ eV, and even in that case the minimum required travel time is just ps or ns for energies of meV or μeV.

### C. Spatial and temporal averaging

Equation (9a) for obtaining a wave function provides a convenient means of handling problems involving all sorts of propagators, even quite complicated ones. However, we still have the potential issue of obtaining many different distributions of paths, depending on $x_{\mathrm{f}}$. For the purpose of displaying a single overall distribution that provides a measure of how a path contributes towards the wave function, we multiply both sides of Eq. (9a) by the complex conjugate of $\psi_j(x_{\mathrm{f}})$ and then to integrate over $x_{\mathrm{f}}$. Applying this procedure, and defining the normalization factor thereby obtained on the left-hand side of Eq. (9a) as

$$N \equiv \int_{\Omega} dx_{\mathrm{f}} \left|\psi_j(x_{\mathrm{f}})\right|^2 , \tag{17a}$$

we then obtain from the right-hand side

$$\frac{1}{N}\int_{\Omega} dx_{\mathrm{f}} \int_{-\infty}^{\infty} dp_{\mathrm{c}} \, \frac{\psi_j^*(x_{\mathrm{f}})\, \Delta\mathcal{J}_j(p_{\mathrm{c}}, x_{\mathrm{f}}, T)}{2\sqrt{\hbar M/T}} = 1 \ . \tag{17b}$$

There are two reasons why this procedure of multiplying by $\psi_j^*(x_{\mathrm{f}})$ and integrating over $x_{\mathrm{f}}$ is suitable for our purpose. First, regarding the phase of the dominant contribution to $\Delta\mathcal{J}_j(p_{\mathrm{c}}, x_{\mathrm{f}}, T)$ (*i.e.* peaks therein), that phase will vary in accordance with the wave function (left-hand side of Eq. (9a)). With the $\psi_j^*(x_{\mathrm{f}})$ multiplier, all peaks will have the same phase (*i.e.* zero), thereby producing a real-valued distribution of paths. Second, if the magnitude of $\psi_j(x_{\mathrm{f}})$ is small, then we need not be so concerned about the paths that compose it. Here again, the $\psi_j^*(x_{\mathrm{f}})$ multiplier provides us with an useful means of forming a weighted average that includes this consideration.

Equation (17) provides a means of describing how specific paths (or sets of paths) as labeled by their $p_{\mathrm{c}}$ values contribute towards forming the eigenstate $j$. We interchange the order of integration in Eq. (17b) and then define

$$\mathcal{P}_j(p_{\mathrm{c}},T) \equiv \frac{1}{N}\int_{\Omega} \frac{\psi_j^*(x_{\mathrm{f}})\,\Delta\mathcal{I}_j(p_{\mathrm{c}},x_{\mathrm{f}})}{2\sqrt{\hbar M/T}}\,dx_{\mathrm{f}} \tag{18a}$$

with the property that

$$\int_{-\infty}^{\infty} \mathcal{P}_j(p_{\mathrm{c}},T)\,dp_{\mathrm{c}} = 1 \tag{18b}$$

which is true for all values of the travel time $T$. Returning to the problem of motion in $\mathbb{R}^1$ discussed in Section II(B), we found in Eq. (15) that $\Delta\mathcal{I}_j(p_{\mathrm{c}},x_{\mathrm{f}},T)/2\sqrt{\hbar M/T} = \exp(ikx_{\mathrm{f}})\,\delta(p_{\mathrm{c}}-\hbar k)$ for $T\to\infty$. The prefactor of $\exp(ikx_{\mathrm{f}})$ goes away when we integrate over final locations as in Eq. (18a), hence yielding $\mathcal{P}_k(p_{\mathrm{c}},T) = \delta(p_{\mathrm{c}}-\hbar k)$. We thus find a distribution with a real, non-negative measure of the contribution that paths make towards forming the wave function. The same distribution is found for motion in $S^1$, whereas two delta functions are produced for reflection from a hard wall and for motion in an infinite square-well.

However, for a harmonic oscillator, it turns out that the distribution $\mathcal{P}_n(p_{\mathrm{c}},T)$ does not have an asymptotic, long-time limit; rather, it is oscillatory (for all values of $T$). This behavior arises from a well-known "focusing" effect in which a particle following any classical trajectory returns to its starting position for times that are multiple of the period of the harmonic motion [16,31,33]. Hence, the contribution that specific paths make towards the eigenstate depends on the travel time. Nevertheless, to gain an overall measure of how a path contributes towards forming an eigenstate, we can average over a period of the motion, evaluating $\langle\mathcal{P}_n(p_{\mathrm{c}},T)\rangle$ in the $T\to\infty$ limit.

As shown in Eq. (18b), real values of $\mathcal{P}_j(p_{\mathrm{c}},T)$ integrate to 1 whereas its imaginary parts integrate to 0. However, in addition, we find that for all the problems analyzed here, the imaginary part of $\mathcal{P}_j(p_{\mathrm{c}},T)$ is 0 (or can be taken to be 0) for *all values* of $p_{\mathrm{c}}$. That is to say, for motion in $\mathbb{R}^1$ as discussed in Section II(B) as well as for all of the problems discussed in Section III we find that the $\mathcal{P}_j(p_{\mathrm{c}},T)$ distributions in the $T\to\infty$ limit can all be taken to be delta functions. For the case of a harmonic oscillator as we discuss in Section IV, its $\mathcal{P}_n(p_{\mathrm{c}},T)$ distributions are also found to be real and non-negative. Thus, for all of these problems, we achieve a rigorous and convenient measure for assessing how paths contribute towards the wave function of a given eigenstate.

## III. Path distributions for simple 1-dimensional problems

### A. Motion around a circle

The situation for free-particle motion around a circle ($S^1$) has been discussed by many authors [3,4,16,22,23,24], but nonetheless we summarize it here as well in order to enable comparison with what occurs for reflection from a hard wall or motion in an infinite square-well. For motion in $S^1$, it is well-known that there are topologically distinct ways for a particle to reach any given final position $\phi_{\mathrm{f}}$ from an initial position $\phi_0$. As illustrated in Fig. 4, the final position can be reached by going around the circle any number of times. The total angle traveled is given by $\phi_n = \phi_{\mathrm{f}} - \phi_0 + 2\pi n$, where the integer $n$ is the winding number. Hence, there are multiple classical trajectories for travel from the initial to the final positions. The $S^1$ propagator contains a term for each such semiclassical path, thus appearing as

$$K^{(S^1)}(\phi_0, \phi_{\mathrm{f}}, T) = \sqrt{\frac{I}{2\pi i\hbar T}} \sum_{n=-\infty}^{\infty} \exp\left(\frac{iI(\phi_{\mathrm{f}} - \phi_0 + 2\pi n)^2}{2\hbar T}\right) \tag{19}$$

where $I = MR^2$ is the moment of inertia of a particle with mass $M$ moving around the circle which has radius $R$. We also note that an additional phase factor (a homotopy factor) can be included in each term of Eq. (19), as discussed by Schulman [3,23], but this factor is not important for the present work so we omit it.

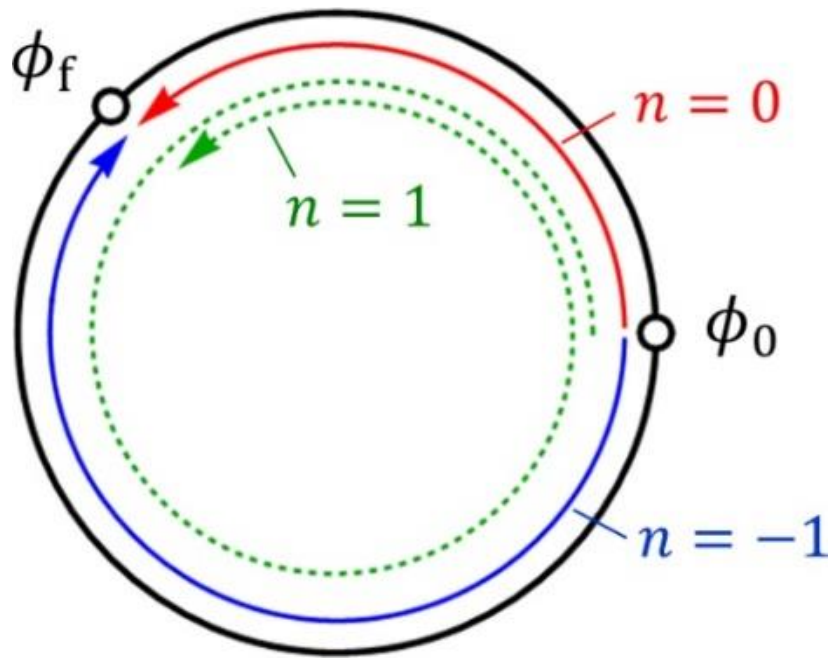

FIG 4. Schematic view of winding numbers for various paths on a circle, with initial angular position $\phi_0$ and final position $\phi_{\mathrm{f}}$. For a given winding number, $n$, the angular distance traveled is $(\phi_{\mathrm{f}} - \phi_0 + 2\pi n)$.

To determine dominant paths, we utilize Eq. (3) with wave function $e^{i\ell\phi}/\sqrt{2\pi}$ having energy eigenvalues $E_\ell = \ell^2\hbar^2/2I$, where $\ell$ is an integer quantum number with $-\infty < \ell < \infty$, thus yielding

$$e^{i\ell\phi_{\mathrm{f}}} = e^{i\hbar\ell^2 T/2I} \int_0^{2\pi} d\phi_0 \; e^{i\ell\phi_0} K^{(S^1)}(\phi_0, \phi_{\mathrm{f}}, T) \tag{20a}$$

$$= e^{i\hbar\ell^2 T/2I} \sqrt{\frac{I}{2\pi i\hbar T}} \int_0^{2\pi} d\phi_0 \; e^{i\ell\phi_0} \sum_{n=-\infty}^{\infty} e^{iI(\phi_{\mathrm{f}} - \phi_0 + 2\pi n)^2/2\hbar T} \tag{20b}$$

$$= e^{i\ell\phi_{\mathrm{f}}} \, e^{i\hbar\ell^2 T/2I} \sqrt{\frac{I}{2\pi i\hbar T}} \sum_{n=-\infty}^{\infty} \int_{\phi_{\mathrm{f}}+2\pi(n-1)}^{\phi_{\mathrm{f}}+2\pi n} d\phi_n \, e^{i2\pi\ell n} \, e^{-i\ell\phi_n} \, e^{iI\phi_n^2/2\hbar T} \tag{20c}$$

$$= e^{i\ell\phi_{\mathrm{f}}} \sqrt{\frac{I}{2\pi i\hbar T}} \int_{-\infty}^{\infty} d\tilde{\phi} \; e^{iI\left(\tilde{\phi} - \hbar\ell T/I\right)^2/2\hbar T} \tag{20d}$$

$$= e^{i\ell\phi_{\mathrm{f}}} \sqrt{\frac{T}{2\pi i\hbar I}} \int_{-\infty}^{\infty} dL_{\mathrm{c}} \; e^{i(L_{\mathrm{c}} - \hbar\ell)^2 T/2\hbar I} \tag{20e}$$

where in Eq. (20c) we rewrite each integral using the variable $\phi_n = \phi_\mathrm{f} - \phi_0 + 2\pi n$. Recognizing that $\exp(i2\pi\ell n) = 1$ for integers $\ell$ and $n$, the sum and integral in Eq. (20c) can then be combined to yield Eq. (20d). We then label the integration variable as simply $\tilde{\phi}$, with extended range of $-\infty < \tilde{\phi} < \infty$ (this variable is the same as the $\phi_n$ of Eq. (20c), but in Eq. (20d) a subscript $n$ on the integration variable would be totally superfluous), and we also complete the square in the exponent there. Finally, in Eq. (20e) we re-express the integral in terms of a characteristic angular momentum

$$L_\mathrm{c} = \frac{I\tilde{\phi}}{T} \ . \tag{21}$$

Stationary-phase analysis of the integral in Eq. (20e) proceeds in exactly the same manner as for the case of motion in $\mathbb{R}^1$. We thus find that in the limit of long-time travel, the motion in an eigenstate for $S^1$ can be described in terms of a *single* type of dominant path, one with $L_\mathrm{c} = \hbar\ell$.

### B. Reflection from a hard wall

It is informative to compare motion in $S^1$ with what occurs when a particle reflects from a hard wall (W) or moves in an infinite square-well. For the former case, with potential of

$$V(x) = \begin{cases} 0 & x > 0 \quad (22a) \\ \infty & x \leq 0 \ , \quad (22b) \end{cases}$$

the propagator is given by [17]

$$K^{(\mathrm{W})}(x_0, x_\mathrm{f}, T) = \sqrt{\frac{M}{2\pi i\hbar T}} \ \left[e^{iM(x_\mathrm{f}-x_0)^2/2\hbar T} - e^{iM(x_\mathrm{f}+x_0)^2/2\hbar T}\right] . \tag{23}$$

This propagator contains multiple (two) terms, similar to the $S^1$ propagator of Eq. (19), but for Eq. (23) there is no distinct topological difference that occurs for the paths associated with the two terms of the propagator [16]. Rather, the terms arise due to the requirement that the propagator, in a time-sliced sense, is the sum of probability amplitudes for all possible paths that uniformly fill the space $x > 0$ [25,26]. This requirement also gives rise to the fact that the second term in Eq. (23) is summed up with a opposite sign relative to the first term, and that in the exponent of the second term the initial position appears with an opposite sign compared to the first term. (Furthermore, this construction yields $\psi(0) = 0$ for the wave function, in accordance with the boundary condition of the Schrödinger equation [17]). Comparing the situation to what occurs for the $S^1$ propagator, a homotopic relationship whereby the paths of this half-space problem can be employed to form the paths of the full $\mathbb{R}^1$ space is not as easy to state here [26]. In any case, so far as our methodology for determining path distributions is concerned, the task for all problems is to deduce a sequence of steps analogous to those of Eq. (20) that produce a final integral which extends between $\pm\infty$.

We substitute Eq. (23) into Eq. (3), utilizing wave functions of $\sin(kx)$ with eigenvalues of $E_k = \hbar^2 k^2/2M$ where $k \geq 0$ ($k$ and $-k$ yield the same wave function, so only one sign for $k$ should be included) and with quantum value of momentum of $\hbar k$. We then rewrite that integral into one that extends over $\pm\infty$,

$$\sin(kx_{\rm f}) = e^{i\hbar k^2 T/2M} \int_0^\infty dx_0 \sin(kx_0)\, K^{(\rm W)}(x_0, x_{\rm f}, T) \tag{24a}$$

$$= \sqrt{\frac{M}{2\pi i\hbar T}}\, e^{i\hbar k^2 T/2M} \left[ \int_0^\infty dx_0 \sin(kx_0)\, e^{iM(x_{\rm f}-x_0)^2/2\hbar T} - \int_0^\infty dx_0 \sin(kx_0)\, e^{iM(x_{\rm f}+x_0)^2/2\hbar T} \right] \tag{24b}$$

$$= \sqrt{\frac{M}{2\pi i\hbar T}}\, e^{i\hbar k^2 T/2M} \left[ \int_0^\infty dx_0 \sin(kx_0)\, e^{iM(x_{\rm f}-x_0)^2/2\hbar T} + \int_{-\infty}^0 d\tilde{x}_0 \sin(k\tilde{x}_0)\, e^{iM(x_{\rm f}-\tilde{x}_0)^2/2\hbar T} \right] \tag{24c}$$

$$= \sqrt{\frac{M}{2\pi i\hbar T}}\, e^{i\hbar k^2 T/2M} \int_{-\infty}^\infty d\tilde{x}_0 \sin(k\tilde{x}_0)\, e^{iM(x_{\rm f}-\tilde{x}_0)^2/2\hbar T} \tag{24d}$$

$$= \sqrt{\frac{M}{2\pi i\hbar T}}\, \frac{e^{i\hbar k^2 T/2M}}{2i} \int_{-\infty}^\infty d\tilde{x}_0 \left[e^{ik\tilde{x}_0} - e^{-ik\tilde{x}_0}\right] e^{iM(x_{\rm f}-\tilde{x}_0)^2/2\hbar T}\,. \tag{24e}$$

where in Eq. (24c) we express the second integral in terms of $\tilde{x}_0 = -x_0$, and this integral is then combined with the first one to obtain Eq. (24d).

With Eq. (24e), we end up with two integrals (one for each of the terms in the square brackets), each of which has the same form as for motion in $\mathbb{R}^1$. Thus, rewriting the integrals using the characteristic (classical) momentum

$$p_{\rm c} = \frac{M(x_{\rm f} - \tilde{x}_0)}{T} \tag{25}$$

as an integration variable, we obtain

$$\sin(kx_{\rm f}) = \frac{1}{2i}\sqrt{\frac{T}{2\pi\hbar i}}\left[ e^{ikx_{\rm f}} \int_{-\infty}^\infty dp_{\rm c}\, e^{i(p_{\rm c}-\hbar k)^2 T/2\hbar M} - e^{-ikx_{\rm f}} \int_{-\infty}^\infty dp_{\rm c}\, e^{i(p_{\rm c}+\hbar k)^2 T/2\hbar M} \right]. \tag{26}$$

Two stationary points are obtained from the exponents in the integrands of this expression, at $p_{\rm c} = \pm\hbar k$. Hence, there are *two* paths that are dominant in producing the value of the wave function, in the limit of long travel-times. Let us consider $x_{\rm f}$ to be fixed and finite. The dominant path with $p_{\rm c} = -\hbar k$ has a final location of $\tilde{x}_0 = x_{\rm f} + \hbar kT/M$ (which is positive since $x_{\rm f} \geq 0$ and $T \geq 0$). This path can be associated with a particle traveling *directly* to this final location. Similarly, the dominant path with $p_{\rm c} = +\hbar k$ has final location $\tilde{x}_0 = x_{\rm f} - \hbar kT/M$ (which is negative, since we are considering large $T$). This path can be associated a particle that travels *indirectly* to the final location, *i.e.* including a reflection from the wall.

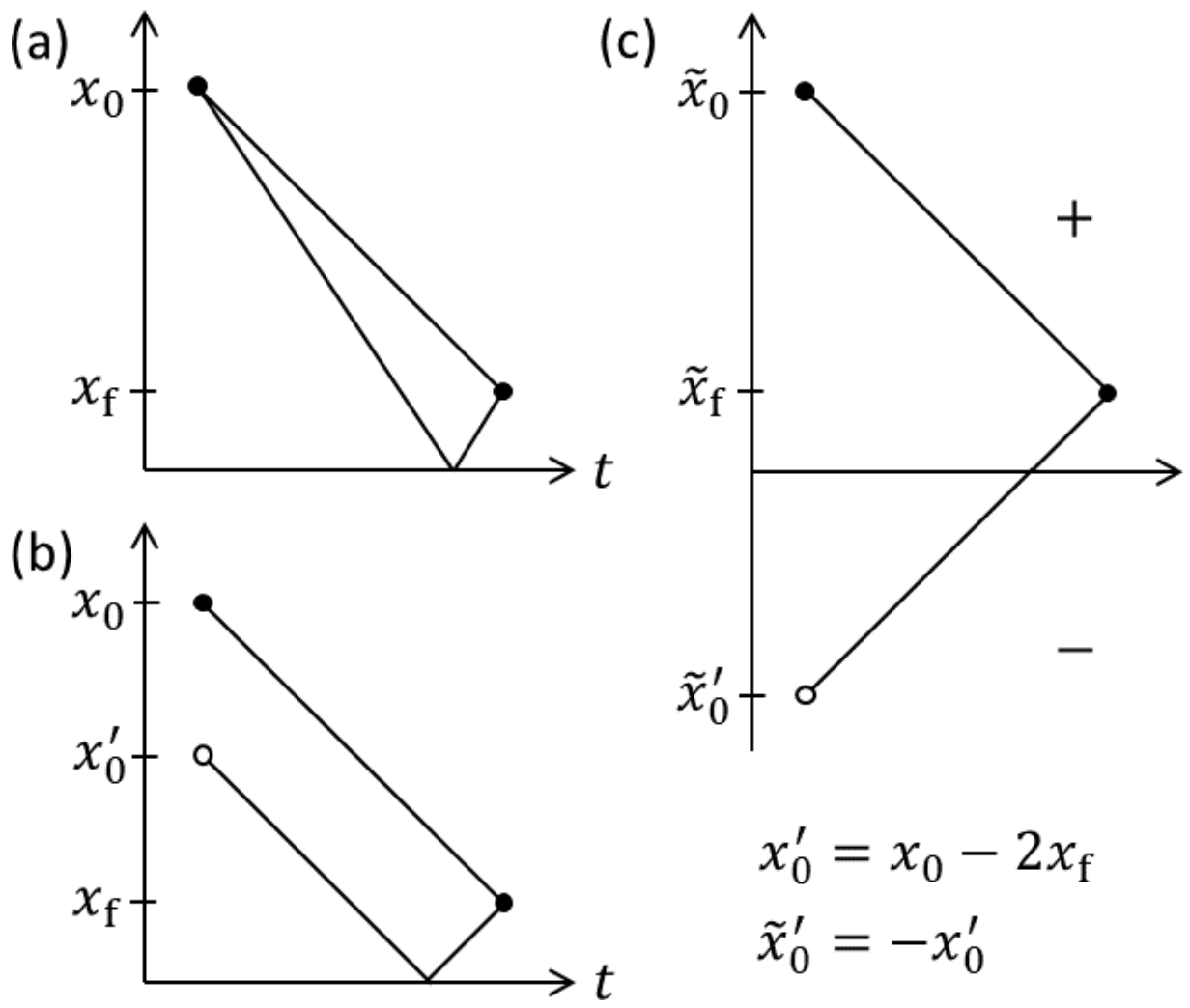

FIG 5. Space-time diagrams of paths for reflection from a hard wall, showing (a) space with $x > 0$ and paths with same initial location but different speeds, (b) space with $x > 0$ and paths with different initial locations but same speed, (c) space $-\infty < \tilde{x} < \infty$ and paths as in (b), with different initial locations but same speed. Expressions for the initial locations $x_0'$ and $\tilde{x}_0'$ of cases (b) and (c) are listed below panel (c), applicable to large travel distance and $x_0 \gg x_{\mathrm{f}}$.

In rewriting the integral of Eq. (24a) defined over $x_0 > 0$ into one that is defined over the extended range of $-\infty < \tilde{x}_0 < \infty$ as in Eq. (24d), and furthermore expressing it in terms of the integration variable $p_{\mathrm{c}}$ of Eq. (25), we have transformed the original problem into an equivalent one. Figure 5 illustrates the various problems we are considering, with Figs. 5(a) and 5(b) both concerning the semi-infinite space of $x > 0$. In Fig. 5(a), paths have the same initial location. The direct path extends to the final location without any reflection from the wall, whereas the indirect path includes a reflection off the wall. In the classical limit of $\hbar \to 0$, these two types of paths would be dominated by straight-line segments as pictured in Fig. 5(a), but the speed ($|p_{\mathrm{c}}|/M$) along the direct and indirect paths would be different.

In contrast, Fig. 5(b) shows direct and indirect paths that leave from different initial locations, but now the travel along these paths has the same speed. This is the situation for the integral in Eq. (26) since is expressed in terms of characteristic momentum. When we consider a limit of $T \to \infty$ for that integral, with fixed momentum (speed), then the limit implies long travel distance. In Fig. 5(b), if we take $x_{\mathrm{f}}$ to have some fixed, finite value, then the $T \to \infty$ limit implies both $x_0 \gg x_{\mathrm{f}}$ and $x_0' \gg x_{\mathrm{f}}$. Figure 5(c) shows the same type of result as in Fig. 5(b), but now drawn using the equivalent problem in which paths can exist over $-\infty < \tilde{x} < \infty$.

Whether or not we view the motion using Fig. 5(b) or 5(c), it is important to note that if we hold the initial location of the direct path of the particle fixed, then the corresponding initial location of the indirect path varies with travel time; this situation is illustrated in Fig. 6. Hence, if we follow a given path as a function of time, then in order to produce the corresponding interfering path we must consider a *new* initial location for that interfering path. Unlike the situation for $\mathbb{R}^1$ in which we can follow any given dominant path as a function of time (increasing $x_{\mathrm{f}}$ while keeping $x_0$ fixed) in order to form the wave function at a later time, as in Fig. 3, this is not possible when we have two dominant paths. In this respect, the situation with two (or more) dominant paths is significantly more difficult to visualize than for a single path.

If we wish to construct color-shaded paths to illustrate the reflection problem, analogous to Fig. 3, the task become increasingly difficult not only due to this need for a varying initial location but also because the wave functions for reflection that do not have any phase variation (*i.e.* they are purely real). Clearly the varying phase of the probability amplitudes *along* each of the respective paths will not be in agreement with the wave function values. It is only *at* the final location of the paths, when the various probability-amplitude terms are summed up (including all prefactors that multiply the probability amplitudes), that a result will be obtained which is in agreement with the wave function. Nevertheless, even with these complications, we can at least conceptually bear in mind a picture somewhat analogous to Fig. 3, with the dominant paths (from various initial locations) extending through space, ending at a final location where their probability amplitudes (including prefactors) sum to yield the wave function.

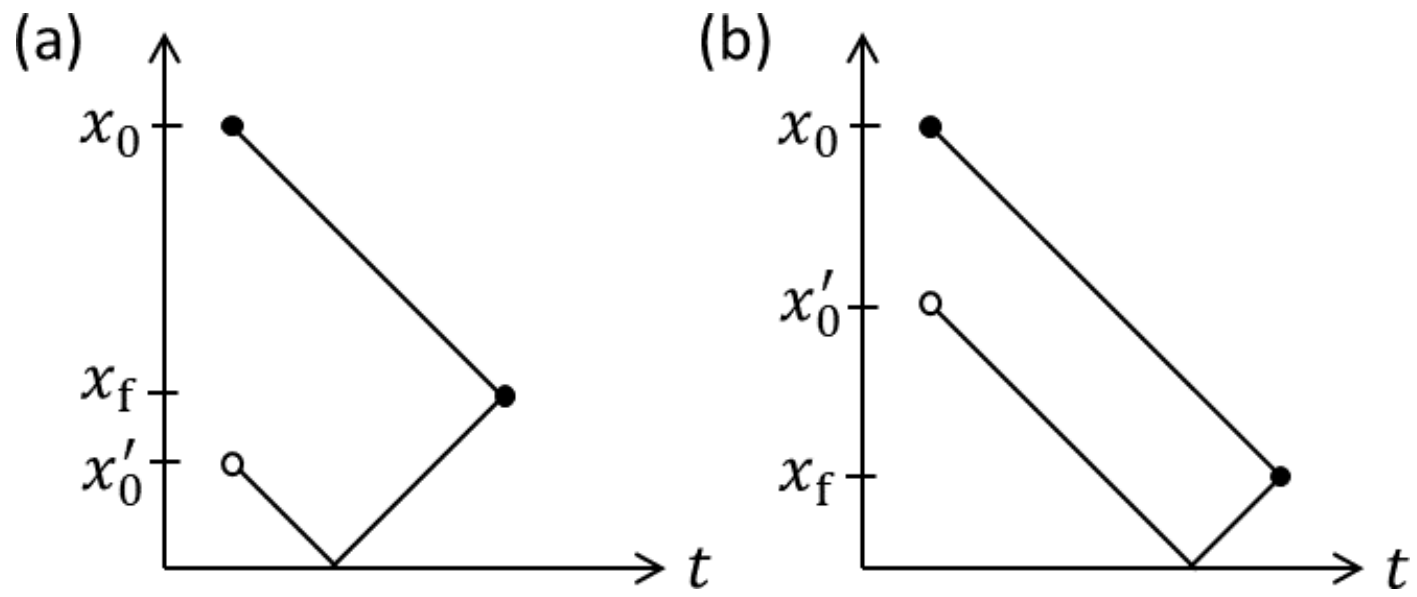

FIG 6. Space-time diagrams of paths for reflection from a hard wall, shown in a manner analogous to that of Fig. 5(b), but now showing two different travel times in (a) and (b).

### C. Motion in an infinite square-well

The propagator for motion in an infinite square-well is well known [16,25,26]; upon insertion into Eq. (3) along with the wave functions and eigenvalues, two changes of variables are needed in order to rewrite the integral: With the square-well defined over $0 < x < a$, a new variable $\hat{x}_0$ with range of $-a \leq \hat{x}_0 \leq a$ is defined, following the same sequence of steps as for the reflection problem of Eq. (24). With that, the problem is transformed into one that is similar to motion on a circle that extends from $-a$ to $+a$, albeit with inclusion of only odd parity solutions (relative to $\hat{x}_0 = 0$) on that circle [30]. That integral is then rewritten following the procedure used for motion around the circle as in Eq. (20), introducing a variable $\tilde{x}_0$ with range $-\infty \leq \tilde{x}_0 \leq \infty$. Further rewriting in terms of the momentum variable $p_\mathrm{c}$ of Eq. (25) then permits stationary-phase analysis. Two stationary points are obtained, in the same way as for the reflection problem, yielding two dominant paths having characteristic momentum of $\pm\hbar k_n$ corresponding to energy eigenvalue for the state being investigated of $\hbar^2 k_n^2/2M$ with $k_n = n\pi/a$ for integer $n \geq 1$.

## IV. Path distributions for the harmonic oscillator

For the harmonic oscillator, the motion ranges over $-\infty < x < \infty$ with potential $M\omega^2 x^2/2$, where $\omega > 0$ is a constant. The classical period of motion is $2\pi/\omega$, and it is a defining feature of the harmonic oscillator that this period is independent of the amplitude of the motion. A closed-form expression for the path integral of the harmonic oscillator is well known [2,31,32,33],

$$K^{(\mathrm{HO})}(x_0, x_\mathrm{f}, T) = \sqrt{\frac{M\omega}{2\pi i\hbar|\sin\omega T|}}\exp\left(i\left[\frac{M\omega\left[\left(x_0^2 + x_\mathrm{f}^2\right)\cos\omega T - 2x_0 x_\mathrm{f}\right]}{2\hbar\sin\omega T} - \frac{\pi}{2}\left\lfloor\frac{\omega T}{\pi}\right\rfloor\right]\right) \quad (27)$$

where $\lfloor u \rfloor$ denotes the floor of $u$, which equals the greatest integer less than or equal to $u$. The term $\lfloor \omega T/\pi \rfloor$ is the Maslov index for the path, corresponding to the number of divergences of $1/|\sin(\omega T)|$ as $t$ varies from 0 to $T$ (or equivalently, the number of full half-periods of the motion) [31,32,33]. Wave functions of eigenstates are $\psi_n(x) = (M\omega/\pi\hbar)^{1/4}\exp(-M\omega x^2/2\hbar)$ $H_n(\sqrt{M\omega/\hbar}\,x)/\sqrt{2^n n!}$ where $H_n$ are Hermite polynomials, with associated eigenvalues of $E_n = (n+1/2)\hbar\omega$ where $n = 0,1,2,....$ Utilizing Eq. (4), the wave function can be written as

$$e^{-M\omega x_\mathrm{f}^2/2\hbar} H_n\left(\sqrt{M\omega/\hbar}\,x_\mathrm{f}\right) = \sqrt{\frac{M\omega}{2\pi i\hbar|\sin\omega T|}}\, e^{i[(n+1/2)\omega T - \pi\lfloor\omega T/\pi\rfloor/2]}$$
$$\times \int_{-\infty}^{\infty} dx_0\, e^{-M\omega x_0^2/2\hbar} H_n\left(\sqrt{M\omega/\hbar}\,x_0\right) \exp\left(\frac{i\omega M\left[\left(x_0^2+x_\mathrm{f}^2\right)\cos\omega T - 2x_0 x_\mathrm{f}\right]}{2\hbar\sin\omega T}\right). \quad (28)$$

The exponent in this integrand describes the action of particle following a classical trajectory extending from $x_0$ to $x_\mathrm{f}$ in travel time $T$.

We rewrite Eq. (28) in terms of a characteristic momentum $p_\mathrm{c}$ that is determined by the endpoints of a path (the paths in this case are all classical trajectories, since the semiclassical form of the propagator is exact). We use for this purpose the *maximum* momentum along a trajectory, corresponding to the momentum when the particle crosses the minimum of the potential. The associated classical energy is $E_\mathrm{c} = (p_\mathrm{c})^2/2M$, and turning points for that motion are given by $\pm p_\mathrm{c}/M\omega$. From the well-known expression for the motion of a trajectory beginning at $x_0$ and ending at $x_\mathrm{f}$ [16],

$$x_\mathrm{c}(t) = x_\mathrm{f}\frac{\sin\omega t}{\sin\omega T} + x_0\frac{\sin\omega(T-t)}{\sin\omega T} \quad (29)$$

where $t$ is the elapsed time along the path, $0 \le t \le T$, the maximum magnitude of the momentum is found to be

$$|p_\mathrm{c}| = \frac{M\omega\left[x_0^2 + x_\mathrm{f}^2 - 2x_0 x_\mathrm{f}\cos\omega T\right]^{1/2}}{|\sin\omega T|}\,. \quad (30)$$

There are two paths for a given $|p_\mathrm{c}|$ that end at the same value of $x_\mathrm{f}$, depending on whether they approach this final position from $x < x_\mathrm{f}$ or $x > x_\mathrm{f}$. These paths have differing values of $x_0$, obtained by solving Eq. (30) for $x_0$

$$x_0 = x_\mathrm{f}\cos\omega T - \mathrm{sgn}(p_\mathrm{c})\sin\omega T\sqrt{(p_\mathrm{c}/M\omega)^2 - x_\mathrm{f}^2} \quad (31)$$

where $\mathrm{sgn}(u)$ is the sign of $u$, which equals 1 for $u > 0$, $-1$ for $u < 0$, or 0 for $u = 0$. Strictly speaking, a quadratic solution of Eq. (30) for $x_0$ would yield a $\pm|\sin\omega T|$ term in Eq. (31) rather than the $-\mathrm{sgn}(p_\mathrm{c})\sin\omega T$ term that is written there, but we are free to use the latter as a particular choice of sign convention for the resulting $x_0$ values. In this way, the value of $x_0$ in Eq. (31) varies with the sign of $p_\mathrm{c}$. Both of the resulting $x_0$ values satisfy Eq. (30). The two paths that are produced approach the final position from $x < x_\mathrm{f}$ or $x > x_\mathrm{f}$, depending on whether $p_\mathrm{c}$ is positive or negative.

We now can rewrite the integral of Eq. (28) in terms of $p_\mathrm{c}$. Utilizing Eq. (31), we find

$$e^{-M\omega x_\mathrm{f}^2/2\hbar} H_n\left(\sqrt{M\omega/\hbar}\,x_\mathrm{f}\right) = \sqrt{\frac{|\sin\omega T|}{2\pi\hbar i M\omega}}\, e^{i[(n+1/2)\omega T - \pi\lfloor\omega T/\pi\rfloor/2]} \int_{-\infty}^{\infty} \frac{|p_\mathrm{c}|\, dp_\mathrm{c}}{\sqrt{p_\mathrm{c}^2 - x_\mathrm{f}^2 M^2\omega^2}}$$

$$\times\ e^{-M\omega x_0^2/2\hbar}\, H_n\left(\sqrt{M\omega/\hbar}\, x_0\right)\, e^{iS/\hbar}\, [\Theta(p_\mathrm{c} - M\omega|x_\mathrm{f}|) + \Theta(-p_\mathrm{c} - M\omega|x_\mathrm{f}|)] \quad (32)$$

where $\Theta$ is a Heavyside step function. Values of $x_0$ here are given by Eq. (31), and we have an action of

$$S = \frac{\sin\omega T}{2M\omega}\left[\cos\omega T\left(p_\mathrm{c}^2 - 2x_\mathrm{f}^2 M^2\omega^2\right) + 2x_\mathrm{f} M\omega\,\mathrm{sgn}(p_\mathrm{c})\sin\omega T\sqrt{p_\mathrm{c}^2 - x_\mathrm{f}^2 M^2\omega^2}\right] \quad (33)$$

as obtained from the exponent on the far right-hand side of Eq. (28), and again using $x_0$ from Eq. (31). Thus, the $\Delta\mathcal{I}_n(p_\mathrm{c}, x_\mathrm{f}, T)$ integral of Eq. (9b) can be formed,

$$\Delta\mathcal{I}_n(p_\mathrm{c}, x_\mathrm{f}, T) = \sqrt{\frac{|\sin\omega T|}{2\pi\hbar i M\omega}}\, e^{i[(n+1/2)\omega T - \pi\lfloor\omega T/\pi\rfloor/2]} \int_{p_\mathrm{c}-\sqrt{\hbar M/T}}^{p_\mathrm{c}+\sqrt{\hbar M/T}} \frac{|p'|\, dp'}{\sqrt{(p')^2 - x_\mathrm{f}^2 M^2\omega^2}} \left(\frac{M\omega}{\pi\hbar}\right)^{1/4}$$

$$\times \frac{e^{-M\omega x_0^2/2\hbar}}{\sqrt{2^n n!}} H_n\left(\sqrt{M\omega/\hbar}\, x_0\right)\, e^{iS/\hbar}\, [\Theta(p' - M\omega|x_\mathrm{f}|) + \Theta(-p' - M\omega|x_\mathrm{f}|)]\,, \quad (34)$$

with $x_0$ again given by Eq. (31) (but using momentum values of $p'$ there).

To obtain an overall characterization of the wave function for a given eigenstate in terms of dominant paths, we average $\Delta\mathcal{I}_n(p_\mathrm{c}, x_\mathrm{f}, T)$ over values of $x_\mathrm{f}$, as discussed in Section II(C). Using Eqs. (18a) and (34), we thus arrive at

$$\mathcal{P}_n(p_\mathrm{c}, T) = \int_{-\infty}^{\infty} \left(\frac{M\omega}{\pi\hbar}\right)^{1/4} \frac{e^{-M\omega x_\mathrm{f}^2/2\hbar}}{\sqrt{2^n n!}} H_n\left(\sqrt{M\omega/\hbar}\, x_\mathrm{f}\right) \frac{\Delta\mathcal{I}_n(p_\mathrm{c}, x_\mathrm{f}, T)}{2\sqrt{\hbar M/T}}\, dx_\mathrm{f}\ . \quad (35)$$

Upon numerical evaluation, we find that $\mathcal{P}_n(p_\mathrm{c}, T)$ is purely real, due to the fact that the imaginary part of $\Delta\mathcal{I}_n(p_\mathrm{c}, x_\mathrm{f}, T)$ turns out to have opposite parity with respect to with inversion of $x_\mathrm{f}$ than does the function $H_n\left(\sqrt{M\omega/\hbar}\, x_\mathrm{f}\right)$. As discussed in Section II(C), Eq. (35) is not amenable to asymptotic analysis for large values of $T$, since it is oscillatory in $T$. This situation is not surprising; it is apparent from the original form of the propagator, Eq. (27), with its $1/\sqrt{|\sin\omega T|}$ divergences. These divergences themselves have been eliminated in Eqs. (32) and (34) due to our use of the $p_\mathrm{c}$ integration variable, but nonetheless the underlying source of the divergences (the focusing effect) still remains and gives rise to the periodicity in those expressions. As detailed in Section II(C), we handle this situation by time-averaging over a period of $2\pi/\omega$, forming

$$\langle\mathcal{P}_n(p_\mathrm{c}, T)\rangle \equiv \frac{\omega}{2\pi}\int_T^{T+2\pi/\omega} \mathcal{P}_n(p_\mathrm{c}, T')\, dT'\ . \quad (36)$$

The integrals in Eqs. (34), (35), and (36) can be easily evaluated numerically, although special attention is required for the integrand in Eq. (34) since it diverges when $p' \to \pm M\omega|x_\mathrm{f}|$. However, the integral as a whole is still finite since this integrand only diverges as $\left[(p')^2 - x_\mathrm{f}^2 M^2\omega^2\right]^{-1/2}$. Thus, very near the divergence, the integral can be evaluated just utilizing the divergent term,

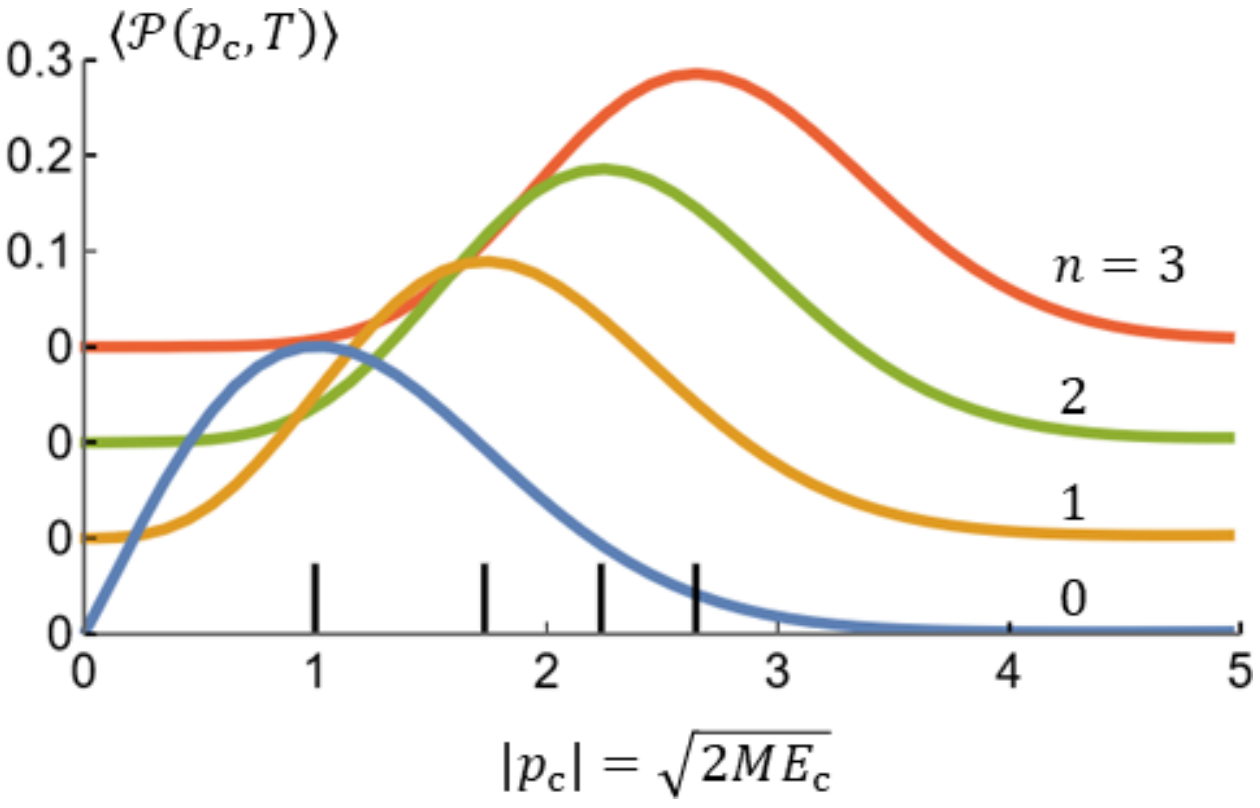

FIG 7. $\langle \mathcal{P}_n(p_c, T) \rangle$ from Eq. (36), as a function of $|p_c|$, for the $n = 0, 1, 2,$ and 3 eigenstates of the harmonic oscillator, evaluated for a travel time of $T = 32\pi$ (using units of $\hbar = M = 1$ and taking $\omega = 1$). The plots are vertically offset from one another, for clarity. The thick lines near the horizontal axis indicate $|p_c|$-values of 1, $\sqrt{3}$, $\sqrt{5}$, and $\sqrt{7}$, corresponding to $\sqrt{2ME_n}$ for the various eigenstates.

$$\int_{M\omega|x_f|}^{M\omega|x_f|+\varepsilon} \frac{|p'|\,dp'}{\sqrt{(p')^2 - M^2\omega^2 x_f^2}} = \int_{-M\omega|x_f|-\varepsilon}^{-M\omega|x_f|} \frac{|p'|\,dp'}{\sqrt{(p')^2 - M^2\omega^2 x_f^2}} = \sqrt{\varepsilon(\varepsilon + 2M\omega|x_f|)} \quad , \tag{37}$$

with $M$, $\omega$, and $\varepsilon$ all being positive, and where all other terms in the integrand of Eq. (34) are evaluated at $p' = \pm M\omega|x_f|$. Other than this special case, we evaluate the integrals in Eqs. (34) – (36) simply using trapezoids to approximate the respective integration areas. With units of $\hbar = M = \omega = 1$, the results shown in Fig. 7 were obtained using $\Delta T = \pi/16$, $\Delta x_f = 0.1$ with $x_f$ ranging between $\pm 5\sqrt{2n+1}$, and $\Delta p' = \left(n_{p'}\sqrt{T}\right)^{-1}$ with $n_{p'}$ being the maximum of 50 or $150|x_f|/\sqrt{2n+1}$. For Fig. 8, $\Delta p_c$ values between 0.01 and 0.1 were used, depending on the plot. With these parameter values, the curves of Figs. 7 and 8 were obtained with approximately one hour of run time each, using a single processor and Mathematica code. The code for Fig. 7 is listed in the Appendix and available at https://www.cmu.edu/physics/stm/publ/136/.

Figure 7 shows results for $\langle \mathcal{P}_n(p_c, T) \rangle$ for $n = 0, 1, 2,$ and 3, computed using a relatively large travel-time of $T = 32\pi$ for which there is little time dependence in the results. Integrating each distribution over the range shown in Fig. 7, $-5 \le p_c \le 5$, yields a value of unity to within 0.015. For each value of $n$, we find that the location of the peak in $\langle \mathcal{P}_n(p_c, T) \rangle$ is equal (within our numerical precision) to what is obtained using $|p_c| = \sqrt{2ME_c}$ and then equating the classical energy $E_c$ to the eigenvalue $E_n$. The results of Fig. 7 are not surprising. It is clear that the harmonic oscillator requires paths over a broad distribution of classical energies ($|p_c|$ values) in order to describe a wave function, since barrier penetration of the wave function necessarily involves paths with classical energy greater than the eigenvalue of energy. Of course, our choice of $p_c$ as corresponding to a *maximum* (signed) value of momentum along the classical trajectories is an arbitrary one; we could have instead chosen *e.g.* the average momentum along a trajectory. Nevertheless, whatever this choice, we find resulting peaks in the distributions for a trajectory having classical energy which is equal to the eigenvalue $E_n$. (Rather than plotting the distributions as a function of $p_c$ with $|p_c| = \sqrt{2ME_c}$ we could instead plot them *vs.* $E_c = p_c^2/2M$ in accordance with

$$\int_{-\infty}^{\infty} dp_c\, \mathcal{P}_n(p_c) = \int_{0}^{\infty} dE_c \sqrt{\frac{2M}{E_c}}\, \mathcal{P}_n\left(\sqrt{2ME_c}\right) \tag{38}$$

with the integrand on the right-hand side used to assess the contribution that paths, labeled by $E_c$, makes towards the given state. However, we find that this distribution with its $1/\sqrt{E_c} = \sqrt{2M}/|p_c|$ weighting factor is not so convenient to utilize since that factor tends to strongly weight paths with small values of $|p_c|$).

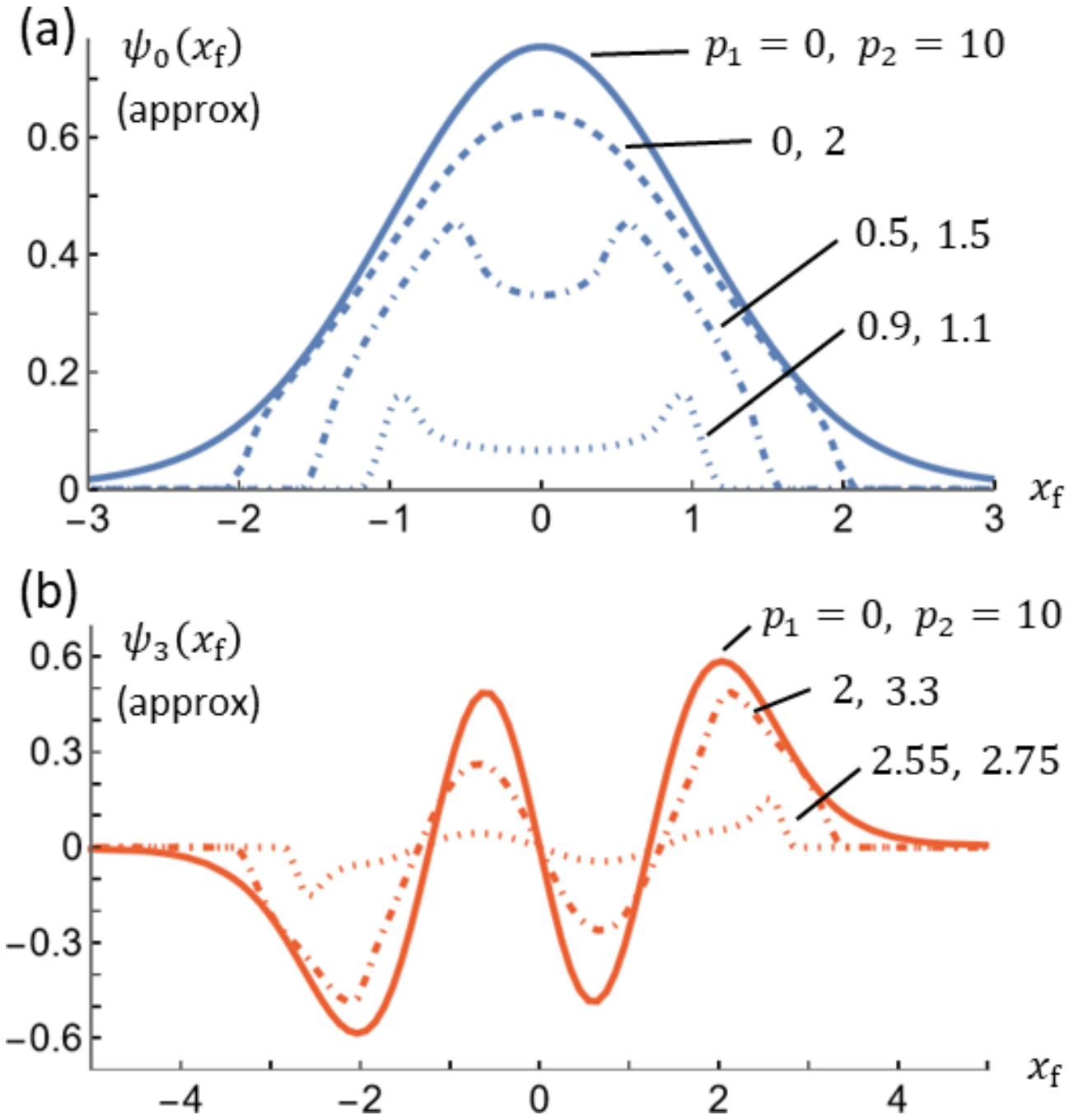

FIG 8. Approximations to the harmonic-oscillator wave function $\psi_n(x_f)$ computed according to Eq. (39) using paths extending over $|p_c| \in (p_1, p_2)$ for the $p_1$ and $p_2$ values indicated. Computations are for a travel time of $T = 32\pi$ (using units of $\hbar = M = 1$ and taking $\omega = 1$). Results are shown for states with (a) $n = 0$ and (b) $n = 3$ (with $\sqrt{2ME_n} = 1$ for (a) and and $\sqrt{2ME_n} = \sqrt{7} \approx 2.65$ for (b)).

To further emphasize the need for a relatively broad distribution of paths needed to describe the eigenstates, in Fig. 8 we investigate formation of wave functions by integrating over paths lying in the range $|p_c| \in (p_1, p_2)$ for various specified values of $p_1$ and $p_2$. Utilizing Eq. (9a), approximations to the wave functions are formed as

$$\frac{\omega}{2\pi} \int_{|p_c| \in (p_1, p_2)} dp_c \int_{T}^{T+2\pi/\omega} \frac{\Delta \mathcal{I}_n(p_c, x_f, T')}{2\sqrt{\hbar M / T'}}\, dT' \tag{39}$$

where we are using the same sort of time average here as in Eq. (36), but now we are *not* integrating over $x_f$ as we did in Eq. (35). The integral over $p_c$ here is restricted to the interval between $p_1 > 0$ and $p_2 > p_1$ for $p_c > 0$, and between $-p_2$ and $-p_1$ for $p_c < 0$. For $p_1 = 0$ and $p_2 \to \infty$, an exact result for $\psi_n(x_f)$ is obtained, in accordance with Eq. (9a). As shown in Fig. 8, for $p_1$ and $p_2$ values that are very near to the $\sqrt{2ME_n}$ peak values seen in Fig. 7, nonzero wave function values only occur in the spatial region $\pm\sqrt{2E_n/M\omega^2}$ (*i.e.* between the turning points that are associated with an energy $E_n$). Again, this is an obvious result – a classical path (extending between the classical turning points) with energy equal to the eigenvalue for the state being examined cannot possibly describe the full wave function. It is only when the range of $p_1$ and $p_2$ is expanded that a realistic form of the wave function is obtained, including tails extending into the classically forbidden region beyond these turning points. It is the paths with $|p| > \sqrt{2ME_n}$ that are responsible for this formation of the wave function tails, while the ones with $|p_c| < \sqrt{2ME_n}$ act to significantly shape the wave function within the classically allowed region of space. The result with $p_1 = 0$ and $p_2 = 10$ shown in Fig. 8 agrees, within the linewidth used for plotting, with the known form of the wave functions as a Hermite polynomial times a Gaussian.

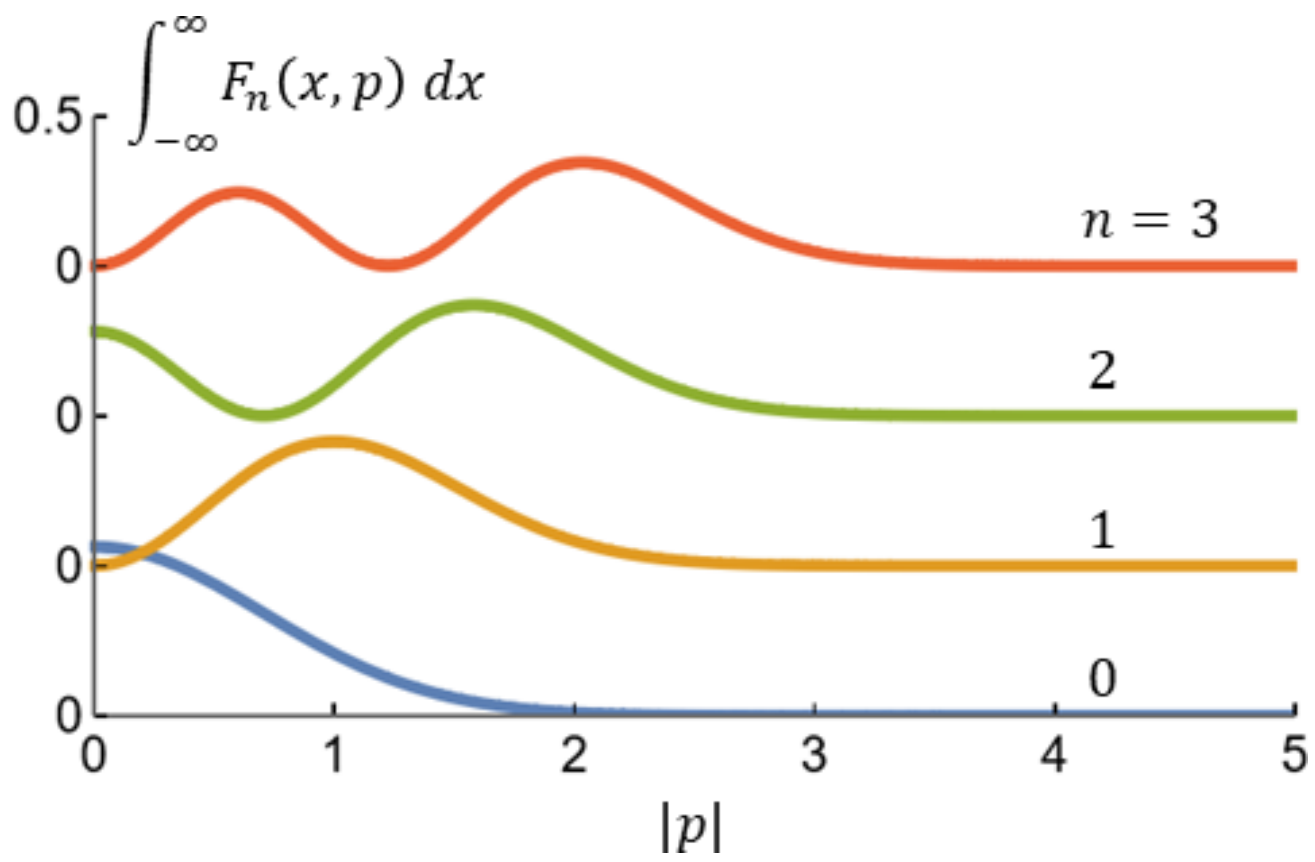

FIG 9. Integrated Wigner distributions, as a function of $|p|$, for the $n = 0, 1, 2,$ and 3 eigenstates of the harmonic oscillator (using units of $\hbar = M = 1$ and taking $\omega = 1$). The plots are vertically offset from one another, for clarity.

## V. Discussion

We point out the difference between the distributions we obtain here for the harmonic oscillator, Fig. 7, and the well-known Wigner distributions. The latter, for the $n^{\text{th}}$ eigenstate of the harmonic oscillator, are given by [34]

$$F_n(x,p) = \frac{(-1)^n}{\pi\hbar} e^{(-M\omega x^2/\hbar - p^2/\hbar M\omega)}\, L_n(2M\omega x^2/\hbar + 2p^2/\hbar M\omega) \qquad (40)$$

where $L_n$ is a Laguerre polynomial of order $n$. These distributions are real-valued, similar to our distributions of Fig. 7, but nevertheless the two types of distributions refer to different variables. Wigner distributions refer to a phase space $(x, p)$ with $p$ being a momentum variable. A Wigner distribution $F_n(x,p)$ integrated over all values of $p$ yields the magnitude squared of the spatial wave function, $|\psi_n(x)|^2$, whereas integrating $F_n(x,p)$ over all values of $x$ yields the magnitude squared of the wave function in momentum space, $\left|\tilde{\psi}_n(p)\right|^2$, where $\tilde{\psi}_n(p)$ is the Fourier transform of $\psi_n(x)$. Hence, we recognize that the momentum variable $p$ in a Wigner distribution refers to *Fourier components* of the wave function, in contrast to the momentum variable $p_c$ that we employ in our path distributions which refers to the action of classical paths of the harmonic oscillator itself (in the present example), *not* those of a free particle. Consequently, if we examine the Wigner distribution integrated over all values of $x$, as in Fig. 9, we see that in general it contains multiple peaks as a function of $p$. Again, these peaks are indicative of the Fourier components of the wave function, *not* the contribution that individual paths make towards the wave function.

This distinction between the variable $p$ as used in a Wigner distribution and the $p_c$ variable that we use for our path distributions is also clearly apparent if we consider motion in an infinite square-well, with zero potential over $0 < x < a$ and infinite potential elsewhere. A common homework problem for students is to compute the momentum-space representation of the associated $\sin(k_n x)/\sqrt{2a}$ wave functions, as given by the associated Fourier transform [35]. Those transforms display peaks at a momentum value of $\pm\hbar k_n$, as expected, but they also have significant sidebands that occur on either side of that peak (since the wave functions are zero for $x$ values outside of $0 < x < a$, thus requiring many Fourier components to produce those zero values). In contrast, our path distributions for this problem as discussed in Section III(C) simply contain two paths, with $p_c = \pm\hbar k_n$. With those, the decomposition of the wave function into $e^{ik_n x}$ and $e^{-ik_n x}$ terms can be easily understood. Surely it is commonplace in a quantum mechanics lecture to speak of a wave function for an infinite square-well as being composed of motion of a particle proceeding to the right and to the left (with the associated terms, *i.e.* probability amplitudes, summing up). The path distributions derived here provide a rigorous means of justifying that picture, since we are dealing with characteristic (classical) paths rather than Fourier components.

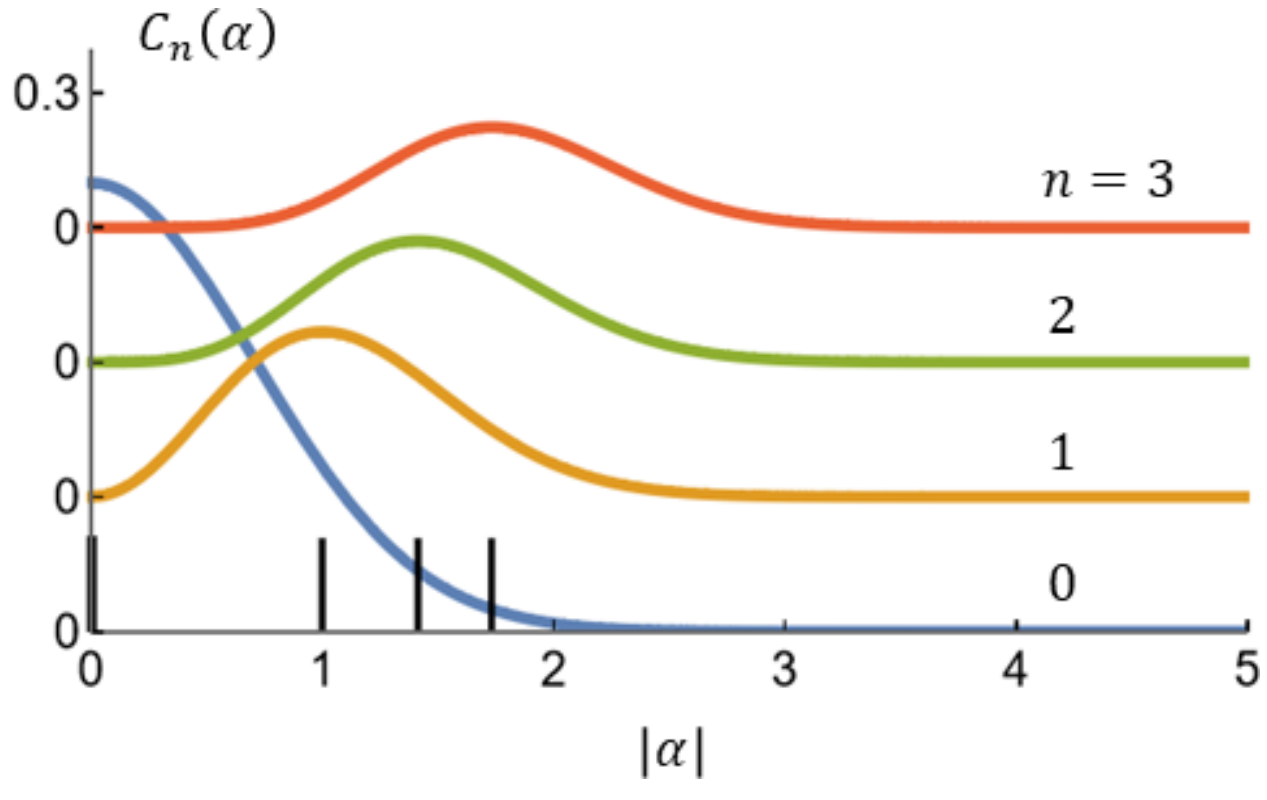

FIG 10. Overlap squared between a harmonic oscillator eigenstate and a coherent state $|\alpha\rangle$, for the $n = 0, 1, 2,$ and 3 eigenstates (using units of $\hbar = M = 1$ and taking $\omega = 1$). The plots are vertically offset from one another, for clarity. The thick lines near the horizontal axis indicate $|\alpha|$-values of $0, 1, \sqrt{2}$, and $\sqrt{3}$, corresponding to expectation values for the energy of the coherent states that equal the energy eigenvalues of the respective eigenstates.

We also compare our path distributions for the harmonic oscillator eigenstates to those obtained from the overlap squared of an eigenstate $|n\rangle$ with a coherent states $|\alpha\rangle$ [36]

$$C_n(\alpha) = |\langle\alpha|n\rangle|^2 = \frac{e^{-|\alpha|^2}\,|\alpha|^{2n}}{n!}\ . \tag{41}$$

This quantity is closely related to a Husimi- or Q-function description of the states [37,38,39,40]. Results for $C_n(\alpha)$ are shown in Fig. 10; they are seen to produce similar behavior as our distributions of Fig. 7, in that there is a single maximum in each function. The expectation value of the energy for a coherent state is $\hbar\omega(|\alpha|^2 + 1/2)$; the corresponding values of $|\alpha|$ when this expectation value equals an energy eigenvalue, $|\alpha| = \sqrt{n}$, are marked by the thick vertical lines in Fig. 10. Just as for Fig. 7, we achieve agreement with the peak position of the corresponding distribution function. In this respect, our path distributions produce similar behavior as the $C_n(\alpha)$ functions. However, the two descriptions are referring to different ways in which to describe the eigenstates, either in terms of position-space paths or using coherent states. It is worth noting in this regard that path integrals themselves can be obtained using either a position-space or a coherent-state formalism [41,42]. The methodology and results of the present work utilize the former, thereby providing an alternate, independent means of obtaining results that are somewhat analogous to those obtained using coherent-state analysis.

Husimi functions, as well as Wigner functions, offer phase-space-based interpretations of quantum mechanics [39,40]. In contrast, for our path distributions, the associated "interpretation" is none other than Feynman's path-based description of quantum phenomena [43]. Our methodology illustrates that path-integral description in ways that go beyond just path distributions, *i.e.* as displayed in diagrams such as Figs. 3 and 8 which involve the use of Eq. (15) *before* it is multiplied by $\psi_j^*(x_\mathrm{f})$ and integrated over $x_\mathrm{f}$ to form the distribution of Eq. (18). As a comparison of Figs. 3(a) and 3(b) makes clear, our treatment is an attempt to illustrate how paths sum up to form a wave function. Of course, path integrals can be discussed with little reference to eigenstates or wave functions (as is indeed the case in Feynman's original description of path integrals [1,2]). Nevertheless, as mentioned in Section I, we *choose* to focus on how paths sum up to form wave functions, due to the familiarity of students with wave functions; our methodology illustrates path integrals in this context, both as an attempt to explain wave functions more fully and as an intended precursor to describing other aspects of path integrals.

Importantly, our results should not be misconstrued as indicating that, somehow, an eigenstate can be understood utilizing a real-valued, non-negative measure for *all possible* paths involved in the motion. On the contrary, we have only demonstrated that if we start from a suitable, closed-form expression of a propagator (such as Eqs. (10), (19), (23), and (27) for motion in $\mathbb{R}^1$, $S^1$, reflection, or harmonic oscillator, respectively), that the paths which are explicitly described by the exponents (actions) of terms in those propagators can be utilized to yield the real-valued, non-negative distributions. However, as mentioned in Section I, the prefactors of the probability-amplitude terms within these propagators do themselves contain the effect of all other paths that differ (fluctuate) from the one that explicitly appears in the exponents (and formation of these fluctuation prefactors [3,4,17,18,19] certainly involve complex-valued sums of probability amplitudes of paths). It is only once the closed-form expression of the propagator is in hand that our methodology can be utilized, yielding a real-valued distribution associated with paths that are specified by the exponents (actions) within the propagator.

We find that this occurrence of real-valued, non-negative measures (in the long time limit) for describing eigenstates is a property of other problems as well. In particular, for motion of a quantum rotator, which is equivalent to motion of a particle on the surface of a sphere ($S^2$), a suitable closed-form expression for the propagator is known [44] and using that we are able to deduce such distributions for that motion [28]. We do not know *why* these real-valued, non-negative distributions are produced for all the situations we have discussed (since, in general, they can be complex-valued), but in any cases it is a very convenient result for the purpose of visualizing motion within the eigenstates.

Moving on to possible application of our path-distribution method for 1-dimensional potentials other than that of the harmonic oscillator, let us consider the "quantum bouncer" with its triangular potential. This problem has been recently been described approximately (but nearly exactly) in terms of a semiclassical propagator [45]. Using that, path distributions for describing eigenstate would be most interesting to evaluate. Perhaps the width or the asymmetry (second or third moments) of these distributions will differ from those of the harmonic oscillator. Perhaps even the peak value (related to the first moment) of the distribution might be shifted slightly away from the energy eigenvalue for the state being investigated. (On a technical note, unlike a time average over one period of the motion that we employed for the harmonic oscillator, the situation for the quantum bouncer or other potentials might require modification of the method since the classical period for the motion depends on energy [45]).

However, notwithstanding these possible application of our path distribution method to barrier-penetration problems in addition to the harmonic oscillator, we feel that it is likely that our method will have limited applicability for problems for which a real-time, semiclassical description does not nearly produce a reasonable description of the dynamics. We mention tunneling problems in particular, for which closed-form expressions for the propagators are known [46,47], but where that propagator has been demonstrated to *not* be amenable to any sort of semiclassical interpretation [47]. For very short travel times, a semiclassical propagator certainly describes the tunneling process (particles travel *over* the barrier) [3,47], but for longer times this expression is invalid. It is not at all clear how to apply our method to this problem, since the known form of the propagator [46,47] cannot be easily rewritten utilizing a signed momentum variable. Further work is required to determine whether path distributions of the type we have developed here can be applied to that sort of problem.

## VI. Summary

In this work we have produced path distributions for a number of problems, utilizing a path-integral method that is a generalization of stationary-phase analysis, applied in the limit of long travel-times. We find distributions that provide a real-valued, non-negative measure of how certain sets of paths, as contained within a closed-form expression of the propagator, contribute towards forming a given wave function. Our distributions are functions of a characteristic momentum $p_\mathrm{c}$ (or angular momentum $L_\mathrm{c}$) that is related to motion between the endpoints of a path for a given travel time. For all of the problems that we have discussed, the characteristic momentum is given simply by $p_\mathrm{c} = \sqrt{2ME_\mathrm{c}}$ (or $L_\mathrm{c} = \sqrt{2IE_\mathrm{c}}$) where $E_\mathrm{c}$ is the energy of a classical trajectory that connects the endpoints of the path. When applied to motion in the simple, 1-dimensional problems of motion in $\mathbb{R}^1$ and in $S^1$, reflection from a hard wall, or motion in an infinite square-well, this method yields one or two delta functions for the resulting path distributions (the actual distributions are slightly more complicated than that, but their effect when used within an integral reduces to simply that of delta functions), with the peak of the delta function occurring at a value of $E_\mathrm{c}$ which equals the energy eigenvalue for the state being considered. For the case of a harmonic oscillator, a broader distribution is obtained, but nevertheless the peak in the distribution still occurs at a classical energy that equals the energy eigenvalue for the state.

## Acknowledgements

The author is very grateful to Riccardo Penco and Michael Widom for numerous discussions. Thanks also go to Debdeep Jena for his comments pertaining to the semiclassical treatment of tunneling, to Xinyu (Alex) Cui for assistance with graphics software, and to both Grayson Frazier and Brandon Neway for a careful reading of the manuscript. This work was supported by the National Science Foundation, grant DMR-1809145.

[25] W. Janke and H. Kleinert, “Summing Paths for a Particle in a Box”, Lett. Nuovo Cimento **25**, 297 (1979).

[26] M. W. Goodman, “Path integral solution to the infinite square well”, Am. J. Phys. **49**, 843 (1981).

[27] J. S. Dowker, “Quantum mechanics and field theory on multiply connected and on homogeneous spaces”, J. Phys. A: Gen. Phys. **5**, 936 (1972).

[28] R. M. Feenstra, “Path distribution for describing eigenstates of orbital angular momentum”, arXiv:2308.02884[quant-ph].

[29] J. S. Townsend (2012), *ibid*., Appendix C.

[30] Kleinert (2009), *ibid.*, Section 6.3.

[31] J.-M. Souriau, ‘‘Construction explicite de l’indice de Maslov. Applications,’’ Lect. Notes Phys. **50**, 125 (1975); *Fourth International Colloqium, Nijmegen*, ed. A. Janner, T. Janssen, and M. Boon (Springer-Verlag, Berlin, 1975).

[32] P. A. Horváthy, “Extended Feynman Formula for Harmonic Oscillator”, Int. J. Theor. Phys. **18**, 245 (1979), and references therein.

[33] P. A. Horváthy, “The Maslov correction in the semiclassical Feynman integral”, Cent. Eur. J. Phys. **9**, 1, (2011).

[34] H.-W Lee, “Theory and application of the quantum phase-space distribution functions”, Physics Reports **259**, 147 (1995).

[35] J. S. Townsend (2012), *ibid.*, problem 6.12, p. 240.

[36] J. S. Townsend (2012), *ibid.*, Chapter 8, Eq. (7.80).

[37] K. Husimi, “Some formal properties of the density matrix”, Proceedings of the Physico-Mathematical Society of Japan **22**, 264 (1940).

[38] N. D. Cartwright, “A non-negative Wigner-type distribution”, Physica **83A**, 210 (1976).

[39] E. Colomés, Z. Zhan, and X. Oriols, “Comparing Wigner, Husimi and Bohmian distributions: which one is a true probability distribution in phase space?”, Journal of Computational Electronics **4**, 894 (2015).

[40] S. Friederich, “Introducing the Q-based interpretation of quantum mechanics”, to appear in British Journal for the Philosophy of Science; arXiv:2106:13502 [quant-ph].

[41] L. S. Schulman (2005), *ibid.*, Chapter 27.

[42] E. Fradkin, *Quantum Field Theory: An integrated approach* (Princeton University Press, Princeton, 2002), Chapter 8 and references therein.

[43] R. P. Feynman, R. B. Leighton, and M. Sands, *The Feynman Lectures on Physics* (Addison-Wesley, Reading, 1965), Volume III, Section 3-1.

[44] R. Camporesi, “Harmonic Analysis and Propagators on Homogeneous Spaces”, Phys. Rep. **196**, 1 (1990).

[45] Y. L. Loh and C. K. Gan, “Path-integral treatment of quantum bouncers”, J. Phys. A: Math. Theor. **54**, 405302 (2021).

[46] M. J. Goovaerts, A. Babcenco, and J. T. Devreese, “A new expansion method in the Feynman path integral formalism: Application to a one-dimensional delta-function potential”, J. Math. Phys. **14**, 554 (1973).

[47] B. Gaveau and L. S. Schulman, “Explicit time-dependent Schrödinger propagators”, J. Phys. A: Math. Gen. **19**, 1833 (1986).

## Appendix

- computations for Fig. 1(a) and Fig. 2
- examine values of R^1 propagator times wavefunction, Eq. (14), as plotted in Fig. 2(a)

```
In[∘]:= Clear["Global`*"];
k = 1; T = 10000; np = 200000; pmin = -6; pmax = 4; delp = (pmax - pmin) / np;
Plot[{Im[Exp[ⅈ (p - k)^2 T / 2] Sqrt[T / (2 π ⅈ)]], Re[Exp[ⅈ (p - k)^2 T / 2] Sqrt[T / (2 π ⅈ)]]},
 {p, 0.7, 1.3}, PlotRange → {{.7, 1.3}, {-70, 70}}, AspectRatio → 1 / 5,
 PlotStyle → {{Thickness[0.008], ColorData[97, "ColorList"]⟦2⟧},
   {Thickness[0.008], ColorData[97, "ColorList"]⟦1⟧}}, AxesOrigin → {0.7, -70}]
```

- examine phasor values (integral of R^1 propagator times wavefunction), Eq. (13), as plotted in Fig. 1(a)

```
In[∘]:= Clear["Global`*"];
k = 1; T = 10000; np = 200000; pmin = -6; pmax = 4; delp = (pmax - pmin) / np;
tab = N[delp (T / (2 π ⅈ))^(1/2)] × Table[
    N[Exp[ⅈ (p - k)^2 T / 2]], {p, pmin, pmax, delp}];
list1 = Accumulate[Re[tab]];
list2 = Accumulate[Im[tab]];
listplot = Transpose[{list1, list2}];
ListLinePlot[listplot, Joined → True,
 AspectRatio → 1 / 2, PlotRange → {{-.5, 1.5}, {-.5, .5}}]
```

- obtain analytic form for integral above (not actually used for plotting)

```
In[∘]:= Clear["Global`*"];
Integrate[(T / (2 π ⅈ))^(1/2) Exp[ⅈ (p - k)^2 T / 2], {p, pc - △p, pc + △p}]
```

Out[∘]=

$$\frac{\left(\frac{1}{2}+\frac{i}{2}\right)\sqrt{T}\left(\mathrm{Erfi}\left[\left(\frac{1}{2}+\frac{i}{2}\right)\sqrt{T}\,(k-pc-\triangle p)\right]-\mathrm{Erfi}\left[\left(\frac{1}{2}+\frac{i}{2}\right)\sqrt{T}\,(k-pc+\triangle p)\right]\right)}{\sqrt{2}\,\sqrt{-i\,T}}$$

- examine the *difference* between phasor values, Eq. (15), as plotted in Fig. 2(b)

```
In[∘]:= nskip = Round[2 / (Sqrt[T] delp)];
Print[nskip];
list1p = Sqrt[T] Drop[list1, nskip] / 2;
list1n = Sqrt[T] Drop[list1, -nskip] / 2;
list1d = list1p - list1n;
list2p = Sqrt[T] Drop[list2, nskip] / 2;
list2n = Sqrt[T] Drop[list2, -nskip] / 2;
list2d = list2p - list2n;
ListLinePlot[{Table[{pmin + (i + nskip / 2) delp, list2d⟦i⟧}, {i, Length[list1d]}],
  Table[{pmin + (i + nskip / 2) delp, list1d⟦i⟧}, {i, Length[list1d]}]},
 PlotRange → {{.7, 1.3}, {-25, 35}}, AspectRatio → 1 / 2,
 PlotStyle → {{Thickness[0.008], ColorData[97, "ColorList"]⟦2⟧},
   {Thickness[0.008], ColorData[97, "ColorList"]⟦1⟧}}, AxesOrigin → {0.7, -25}]
```

- computations for Fig. 7, providing results of Eq. (36) using Eqs. (35), (34), (33) and (31)
- in variable names below, an appended ‘p’ means prime (so Tp = T’ (T-prime), pp = p’ (p-prime), etc.)
- turning points are when M $\omega^2 x^2/2 = (n + 1/2)$ hbar $\omega$, so x = Sqrt[(2n + 1) hbar / M $\omega$]; perform xf integral over 5 of these ranges for either sign of xf
- use specified nTp number of points for the integral over Tp (32 for fine computations, 2 for coarse)
- compute results from p = 0 to 5 with specified delp spacing (0.1 for fine computations, 0.5 for coarse)
- includes correction on first discrete point evaluated in pp integral, near the 1/Sqrt$\left[\text{pp}^2 - \text{xf}^2\ \omega^2\right]$ divergence (integrate that term over the delpp interval)

```
In[∘]:= Clear["Global`*"];
ω = 1;  T = N[32 π];  nTp = 32;  delTp = N[2 π / nTp];
pmax = 5;  delp = 0.1;  np = Round[pmax / delp];
xfmax = N[5 Sqrt[(2 n + 1) / ω]];  delxf = 0.1;
nxf = 2 Ceiling[xfmax / delxf];  delxfp = 2 xfmax / nxf;
npp = Floor[Max[50, 150 Abs[xf] / (Sqrt[(2 n + 1) / ω])]];
prefact[ω_, T_] :=
  N[(ω / N[π])^(1/2) Sqrt[Abs[Sin[T ω]]] / (2^n Factorial[n] Sqrt[2 N[π] ⅈ ω]) Exp[ⅈ ((n + 1 / 2) ω T - π Floor[ω T / N[π]] / 2)]];
x0[p_, xf_, ω_, T_] := xf Cos[T ω] - Sign[p] Sin[T ω] √((p / ω)^2 - xf^2);
S[p_, xf_, ω_, T_] :=
  N[(Sin[T ω] / (2 ω)) (Cos[T ω] (p^2 - 2 xf^2 ω^2) + 2 xf ω Sign[p] Sin[T ω] √(p^2 - xf^2 ω^2))];
tab[n_, xfmax_, nxf_, delxfp_, np_] := Table[{p = (ip + 0.5) delp;  p,
    delxfp Total[Table[xf = -xfmax + (ixf - 0.5) delxfp;
       Exp[-ω xf^2 / 2] HermiteH[n, √ω xf] (2 N[π])^-1 delTp
        Total[Table[Tp = N[T + (iTp - 0.5) delTp];  delpp = 1 / (Sqrt[Tp] npp);
          (Sqrt[Tp] / 2) Re[prefact[ω, Tp] × Total[Table[pp = p + (ipp + 0.5) delpp;  If[
                pp > ω Abs[xf] || pp < -ω Abs[xf], If[(pp > 0 && (pp - ω Abs[xf]) ≤ delpp) ||
                  (pp < 0 && (-ω Abs[xf] - pp) ≤ delpp),
                 Sqrt[(Abs[pp] + 0.5 delpp)^2 - xf^2 ω^2] HermiteH[n, √ω x0[pp, xf, ω, Tp]]
                  Exp[-ω x0[pp, xf, ω, Tp]^2 / 2] Exp[ⅈ S[pp, xf, ω, Tp]],
                 delpp (Abs[pp] / Sqrt[pp^2 - xf^2 ω^2]) HermiteH[n, √ω x0[pp, xf, ω, Tp]]
                  Exp[-ω x0[pp, xf, ω, Tp]^2 / 2] Exp[ⅈ S[pp, xf, ω, Tp]]], 0], {ipp,
                -npp, npp - 1}]]], {iTp, nTp}]], {ixf, nxf}]]}, {ip, 0, np - 1}];
n = 0;
tab0 = tab[n, xfmax, nxf, delxfp, np];
fname = "tab0.dat";  Print[fname];
stmp = OpenWrite[fname];  Write[stmp, tab0];  Close[stmp];
Print["total = ", 2 delp Total[Table[tab0〚ip〛〚2〛, {ip, np}]]];
n = 1;
tab1 = tab[n, xfmax, nxf, delxfp, np];
```

```
fname = "tab1.dat"; Print[fname];
stmp = OpenWrite[fname]; Write[stmp, tab1]; Close[stmp];
Print["total = ", 2 delp Total[Table[tab1[[ip]][[2]], {ip, np}]]];
n = 2;
tab2 = tab[n, xfmax, nxf, delxfp, np];
fname = "tab2.dat"; Print[fname];
stmp = OpenWrite[fname]; Write[stmp, tab2]; Close[stmp];
Print["total = ", 2 delp Total[Table[tab2[[ip]][[2]], {ip, np}]]];
n = 3;
tab3 = tab[n, xfmax, nxf, delxfp, np];
fname = "tab3.dat"; Print[fname];
stmp = OpenWrite[fname]; Write[stmp, tab3]; Close[stmp];
Print["total = ", 2 delp Total[Table[tab3[[ip]][[2]], {ip, np}]]];
```

- output for nTp=32; delp=0.1;

```
tab0.dat
total = 1.00194
tab1.dat
total = 1.00766
tab2.dat
total = 1.00565
tab3.dat
total = 1.01348
```

- output for nTp=2; delp=0.5;

```
tab0.dat
total = 0.991038
tab1.dat
total = 0.99994
tab2.dat
total = 0.974582
tab3.dat
total = 1.00017
```